\documentstyle[flushrt,onecolumn,psfig]{mn}
\newif\ifAMStwofonts 
\AMStwofontstrue 
\title[Truncated isothermal spheres]
	{A model for the postcollapse equilibrium of cosmological structure:
	 truncated 
	isothermal spheres from top-hat density perturbations }
\author[P.R.~Shapiro, I.T.~Iliev and A.C.~Raga]{
Paul R. Shapiro,$^1$ Ilian T. Iliev$^2$ and  Alejandro C. Raga$^3$\\
$^1$Dept. of Astronomy, The University of Texas at Austin,
          Austin, TX 78712, USA, E-mail: shapiro@astro.as.utexas.edu\\
$^2$Dept. of Physics, The University of Texas at Austin,
          Austin, TX 78712, USA, E-mail: iliev@astro.as.utexas.edu\\
$^3$Instituto de Astronom\'\i a-Universidad Nacional Autonoma de 
	M\'exico, Apdo Postal 70-264, 04510 M\'exico D.F., M\'exico,\\ 
	E-mail: raga@astroscu.unam.mx}
\begin{document}
\label{firstpage}
\maketitle
\newcommand{\dl}{\delta_L}
\newcommand{\dc}{\delta_c}
\newcommand{\di}{{\delta_i}}
 
\begin{abstract}
The postcollapse structure of objects which form by gravitational condensation
out of the expanding cosmological background universe is a key element in the
theory of galaxy formation. Towards this end, we have reconsidered 
the outcome of the nonlinear growth of a uniform, spherical density 
perturbation in an unperturbed background universe -- the 
cosmological ``top-hat'' problem.
We adopt the usual assumption that the collapse to infinite 
density at a finite time predicted by the top-hat solution is interrupted by
a rapid virialization caused by the growth of small-scale inhomogeneities in 
the initial perturbation. We replace the standard description of the 
postcollapse object as a uniform 
sphere in virial equilibrium by a more self-consistent one as a truncated,
nonsingular,  
isothermal sphere in virial and hydrostatic equilibrium, including for the 
first time a proper treatment of the finite-pressure boundary condition on 
the sphere. The results differ significantly from both the uniform sphere
and the {\it singular} isothermal sphere approximations for the postcollapse 
objects. The virial temperature which results is more than twice the 
previously used ``standard value'' of the postcollapse uniform sphere 
approximation but $1.4$ times smaller than that of the singular, 
truncated isothermal sphere approximation. 
The truncation radius is 0.554 times the radius of the top-hat at maximum 
expansion, and the ratio of the truncation radius 
to the core radius is 29.4, yielding a central density which is $514$ 
times greater than at the surface and $1.8\times10^4$
times greater than that of the unperturbed background density at the epoch 
of infinite  collapse predicted by the top-hat solution. For the 
top-hat fractional overdensity $\dl$ predicted by extrapolating the linear 
solution into the nonlinear regime, the standard top-hat model assumes that
virialization is instantaneous at $\dl=\dc=1.686$, i.e. the epoch at which
the nonlinear top-hat reaches infinite density. The surface of the 
collapsing sphere meets that of the postcollapse equilibrium sphere
slightly earlier, however, when $\dl=1.52$. These results will have a 
significant effect on a wide range of applications of the Press-Schechter and
 other semi-analytical models to cosmology. 

\vskip10pt

We discuss the density profiles obtained here in relation to the density 
profiles for a range of cosmic structures, from dwarf galaxies to galaxy 
clusters, indicated by observation and by N-body simulation of cosmological 
structure formation, including the recent suggestion of a universal density 
profile for halos in the Cold Dark Matter (CDM) model. The nonsingular 
isothermal 
sphere solution presented here predicts the virial temperature and integrated 
mass distribution of the X-ray clusters formed in the CDM model as found by
detailed, 3D, numerical gas and N-body dynamical simulations remarkably well.
This solution allows us to derive analytically the numerically-calibrated 
mass-temperature and radius-temperature scaling laws for X-ray clusters which 
were derived empirically by Evrard, Metzler and Navarro from simulation results
for the CDM model.
\end{abstract}

\section{Introduction}

The theory of structure formation in cosmology must answer the question of 
how gravitational instability amplified initially small-amplitude density
fluctuations in an expanding background universe until the nonlinear structure 
we see today, from galaxies to clusters of galaxies, emerged. It must explain 
the number and frequency of these objects, their spatial clustering and 
relative motions, their internal structure, and the evolution of these 
properties. The current view is that this structure arose by the complex, 
hierarchical build-up which resulted from a primordial spectrum of 
Gaussian-random noise which reached nonlinear amplitude some time after matter
began to dominate the total energy density of the universe. The exact nature of
this structure formation is not amenable to an analytical solution. As a
result, much effort has gone into large-scale numerical simulation by N-body
techniques and, more recently, the coupling of N-body and hydrodynamics 
methods. For many purposes, however, analytical and semi-analytical 
approximations are possible and serve both as a guide to understanding the 
complex numerical results and as a crucial means of extrapolating beyond them 
in order to interpret the predictions of cosmological models in comparison 
with the observed universe. 

The nonlinear evolution of cosmic density fluctuations is often approximated 
by a model in which the initial linear perturbation is an isolated, uniform 
sphere outside of which the matter is unperturbed---the ``top-hat'' 
perturbation --- for which an exact solution is possible (e.g. 
Peebles 1980, Peebles 1993,  Padmanabhan 1993; and references therein). 
For many purposes, a critical part of the application of this 
top-hat model involves the assumption that the collapse to infinite density 
predicted by the solution is interrupted by a rapid relaxation to virial 
equilibrium at a finite density, as a result of the growth of initially small
inhomogeneities in the density distribution. For example, many applications of
the Press-Schechter approximation \cite{PS} for the rate of formation of 
objects of a given mass per unit time due to the growth
 of Gaussian-random-noise density perturbations involve this assumption of 
postcollapse virialization (see, for example, Monaco 1997 for a review and 
references). 

We have reconsidered the postcollapse virial equilibrium of the top-hat 
model. We improve upon the standard 
assumption of a uniform sphere for the virialized final state by a 
self-consistent model involving a nonsingular, truncated isothermal sphere 
(``TIS'') in
hydrostatic and virial equilibrium which takes proper account of the 
finite-pressure boundary condition on the sphere. The results differ 
significantly from those of the standard uniform sphere (``SUS'') approximation
for the final virialized object. They also differ significantly from 
those of the approximation of a {\it singular} isothermal sphere (``SIS'')
for the final object.

An outline of the paper is as follows. In \S~\ref{top} we describe the 
top-hat model and the standard approach to the final virialized object in the
SUS approximation. In \S~\ref{isoth} we describe the hydrostatic equilibrium 
solutions for isothermal spheres. An application of the virial theorem to 
characterize the equilibrium of a truncated isothermal sphere, including a
proper account of the finite boundary pressure, is described in 
\S~\ref{virial}. A minimum-energy argument which selects a unique TIS solution
for a given boundary pressure is presented in \S~\ref{min_en}. This solution 
is used to specify the postcollapse TIS which results from any given top-hat
density perturbation in \S~\ref{TIS}. An independent calculation of this
postcollapse TIS solution is presented in \S~\ref{Bert}, based on the 
self-similar infall solution of Bertschinger \shortcite{Bert85}, with 
results which agree closely with the minimum-energy argument, thereby 
confirming it. Our results are summarized and discussed in \S~\ref{results}.
Our summary in \S~\ref{summary_tis} includes a complete prescription, with
scaling laws and an approximate fitting formula, for using our minimum-energy 
TIS solution in combination with the top-hat perturbation solution to
describe the cosmological formation and postcollapse equilibrium structure of
any object as a function of its mass and collapse epoch. In  
\S~\ref{compare_SUS_SIS}, we compare the predictions of this TIS solution 
with those of the SUS and SIS approximations. A comparison of the TIS density 
profile with existing results of numerical simulations of cosmic structure
formation and with observations for a range of cosmic structures from
dwarf galaxies to galaxy clusters is made in \S~\ref{compare_dens}. This
includes a discussion of the minimum-energy TIS solution in relation to the
suggestion by Navarro, Frenk, and White (1996, 1997) of a universal form 
for the density profiles of haloes of collisionless dark matter formed in a
hierarchical clustering model like the Cold Dark Matter (CDM) model, based upon
the results of numerical N-body simulation. Finally, in  \S~\ref{compare_temp},
we compare our predicted virial temperature with the results of numerical gas
dynamical simulations of X-ray cluster formation in the CDM model, and
derive analytically the mass-radius-temperature scaling laws for X-ray clusters
determined empirically by Evrard, Metzler, and Navarro (1996) from gas 
dynamical and N-body simulations of the CDM model.

%
%
\section{The standard Approach to the Top-Hat model}
\label{top}
All the results in this section are well known, but due to their importance 
in what follows we shall summarize  them here.

\subsection{Before collapse: the exact nonlinear solution}
The spherical top-hat model, a spherical perturbation of uniform overdensity
\cite{G&G}, affords considerable insight into the 
dynamics of gravitational growth of cosmic structure, while still having an 
exact, analytical or semi-analytical, solution in certain cosmological models
(cf. Peebles 1980). In what follows we shall consider top-hat perturbations
only in the Einstein-de Sitter model.

Based on its high degree of symmetry, the collapse 
of a top-hat density perturbation can be described in detail solely in
terms of its overdensity with respect to the background. From the Birkhoff
theorem, the equation
describing this evolution is identical to the Friedmann equation for a
universe with the corresponding mean density. The linearized version of this 
equation describes the linear evolution of the density perturbation
$\delta_L(t)$ and has a growing solution $\delta_L=\di(t/t_i)^{2/3}$. Using 
this solution we can write the equation for the density contrast in terms of
$\delta_L(t)$, whose solution is independent
of the initial density contrast $\di$ and the corresponding time $t_i$ and
hence describes the general, scale-invariant collapse solution. 
In parametric form the solution reads (e.g. 
Padmanabhan 1993)
\begin{equation}
\begin{array}{ll}
\label{exact}
\displaystyle{\delta=
	\frac 92 \frac{(\theta-\sin\theta)^2}{(1-\cos\theta)^3}-1},\qquad
\displaystyle{\delta_L
=\frac 35 \left(\frac34\right)^{2/3} (\theta-\sin\theta)^{2/3}} \,.
		\end{array}
\end{equation}
The critical density contrast $\dc$ is defined as the linear solution 
extrapolated to  the epoch at which the nonlinear solution predicts an 
infinite density. As described above, this effectively is the time of 
collapse in scale-free units. The values obtained at turnaround 
($\theta=\pi$, indicated by subscript ``$m$'' to denote ``maximum expansion'')
 and at collapse time ($\theta=2\pi$,
 denoted by subscript $c$) are:
\begin{equation}
\begin{array}{ll}
\label{deltas}
\delta_{Lm}=\displaystyle{\frac35\left(\frac{3\pi}4\right)^{2/3}}
	\approx1.0624 \,,
	\qquad\delta_m=\displaystyle{\frac{9\pi^2}{16}-1\approx 4.5517},
	\nonumber \qquad
\delta_c=\displaystyle{\frac35\left(\frac{3\pi}2\right)^{2/3}}
	\approx1.6865 \,, 
	\qquad\delta(\delta_c)=\infty. 
\end{array}
\end{equation}
Henceforth, we shall refer to $z_{\rm coll}$ as the redshift which corresponds
to the epoch of infinite collapse, at which $\delta=\infty$, at time 
$t_{\rm coll}$.
\subsection{After collapse: uniform sphere in virial equilibrium}

As shown above, a perfectly symmetric top-hat collapse results in a singularity
at $\delta_L=\dc$, which does not lead to the formation of virialized 
structures.
To overcome this problem, it is usually assumed that the actual collapse is 
slightly inhomogeneous, and therefore the top-hat does not collapse to 
infinite density, but instead, by means of processes like violent 
relaxation, relaxes to form a static, virialized structure. Assuming that 
the total energy is conserved during the collapse, we can connect the 
initial top-hat to the final state as follows.
 
The standard approach is to assume that the collapse of the top-hat
to infinite density is interrupted by a rapid equilibration at $\dl=\dc$
which results in another uniform sphere in virial equilibrium.
We solve for the final radius $r_{\rm vir}$ of the virialized sphere by applying
the virial theorem to the final state and expressing the conserved total energy
$E$ of the sphere in terms of the radius $r_m$ at maximum expansion.
At the point of maximum expansion (which always exists for $\Omega\geq 1$) the
sphere is cold and at rest, so its energy is entirely gravitational potential
energy which, for uniform sphere of mass $M_0$ and radius $r_m$ is just
\begin{equation}
\label{en1}
E = W_m= - \frac 35 \displaystyle{\frac {GM_0^2}{r_m}}.
\end{equation}
After the system settles down to a virial equilibrium 
its potential energy is 
\begin{equation}
\label{old_pot_en}
W_{\rm vir} = - \frac 35 \displaystyle{\frac {GM_0^2}{r_{\rm vir}}}.
\end{equation}
According to the virial theorem 
\begin{equation}
0=2K+W,
\end{equation}
where $K=U_{\rm th}+T_{\rm kin}$, $U_{\rm th}$ and $T_{\rm kin}$ are the 
thermal and kinetic energy, respectively.
Together with the conservation of energy, $E=K+W$, this implies $E=-W/2$,
and, therefore, the total energy is related to the virial radius according to
\begin{equation}
\label{en2}
E = - \frac 3{10} \displaystyle{\frac {GM_0^2}{r_{\rm vir}}}.
\end{equation}
Equating the total energy $E$ in equations (\ref{en1}) and (\ref{en2}) yields
\begin{equation}
r_{\rm vir}=\frac{r_m}2.
\end{equation}
This is the well-known result for the size of the virialized object in the SUS
 approximation. The virial temperature and velocity are derived as follows.
The kinetic energy in the virialized state is 
\begin{equation}
K_{\rm vir}=E-W_{\rm vir} 
	= \frac 3{10} \displaystyle{\frac {GM_0^2}{r_{\rm vir}}}.
\end{equation}
This kinetic energy is the energy of internal motions only ($T_{kin}=0$).
To obtain the equivalent temperature we express the thermal energy
\begin{equation}
U_{\rm th}=\frac 32\displaystyle{ \frac{k_B T}{m}M_0},
\end{equation}
where $m$ is the mean mass per gas particle. If $m_H$ is the mass of a hydrogen 
atom, then $m=\mu m_H$ where $\mu$ is the mean molecular weight.
Therefore the equivalent virial temperature is
\begin{equation}
\label{temp}
T =\displaystyle{\frac 1 5\frac {GM_0m}{k_Br_{\rm vir}}}
=\displaystyle{\frac 2 5\frac {GM_0m}{k_Br_m}}.
\end{equation}
Henceforth, we shall refer to the virial temperature in equation (\ref{temp}) 
as $T_{SUS}$. For the case of a collisionless gas, we replace the virial 
temperature above by the virial velocity dispersion,
\begin{equation}
\label{dispersion}
\sigma^2=\frac{\langle v^2\rangle}3=\frac{k_BT}{m}
\end{equation}

According to this approach, the properties of the final virialized uniform 
sphere are related to those of the top-hat perturbation which created it as 
follows, as summarized by Padmanabhan\shortcite{Padmanabhan}, except we use
$h=H_0/(100 {\rm\, km\,s^{-1}\,Mpc^{-1}})$ rather than $h_{0.5}=2h$, and we 
include the explicit dependence on the mean molecular weight $\mu$, where
$\mu= 0.59\, (1.22)$ for an ionized (neutral) gas of H and He with 
$[He]/[H]=0.08$ by number:
\begin{equation}
r_m = 338.3 \left(\frac M{10^{12}M_{\odot}}\right)^{1/3}(1+z_{\rm coll})^{-1}
		h^{-2/3} {\rm kpc},
\end{equation}
\begin{equation} 
(r_{\rm vir})_{SUS} 
    = 169.2 \left(\frac M{10^{12}M_{\odot}}\right)^{1/3}
	(1+z_{\rm coll})^{-1}h^{-2/3} {\rm kpc},
\end{equation}
\begin{equation}
T_{SUS}=6.160\times 10^5\mu\left(\frac M{10^{12}M_{\odot}}\right)^{2/3}
	(1+z_{\rm coll})h^{2/3} {\rm K},
\end{equation}
\begin{equation}
(v_c)_{SUS}=5^{1/2}\sigma=\left(\frac53\right)^{1/2}\langle v^2\rangle^{1/2}
	= 159.4\left(\frac M{10^{12}M_{\odot}}\right)^{1/3}
	(1+z_{\rm coll})^{1/2}h^{1/3} {\rm km\, s^{-1}},
\end{equation}
\begin{equation}
(\bar{\rho})_{SUS}= 18\pi^2(1+z_{\rm coll})^3\rho_{b0},
\end{equation}
where $v_c=(GM/r_{\rm vir})^{1/2}$ is the circular velocity at $r_{\rm vir}$. 

This approach is commonly used to calculate the virial temperature and 
radius. However it is not a realistic model, since it assumes isothermality, 
uniform density, and the absence of external pressure,
which are clearly incompatible. A more realistic 
approach is to assume a final state of 
hydrostatic equilibrium (or the corresponding state 
with an isotropic Maxwellian velocity distribution in the case of collisionless 
particles), which we will do in the next section.

%
%
\section{Isothermal Spheres}
\label{isoth}
The final virialized object is decoupled from the expanding cosmological 
background from which it condensed. Hence, when we describe it as an 
isothermal sphere in hydrostatic equilibrium, we do so in the usual 
non-cosmological way (cf. Binney and 
Tremaine 1987).
The hydrostatic equilibrium equation, $\nabla p = \rho {\bf g}$,
in the case of spherical symmetry becomes
\begin{equation}
\label{hydrost_sph}
\displaystyle{T\frac{k_B}m \frac{d\rho}{dr} 
	= - \rho g =-\rho \frac{GM(r)}{r^2}}
\end{equation}
where $M(r)$ is the mass inside radius $r$. 
Multiplying equation (\ref{hydrost_sph}) by $r^2m/\rho k_B T$, and taking the 
derivative with respect to $r$, we obtain
\begin{equation}
\label{eqrho_sph}
\displaystyle{\frac{d}{dr} \left(r^2 \frac{d(\ln \rho)}{dr}\right)
	= -4 \pi \frac{Gm}{k_B T} r^2 \rho}.
\end{equation}

Let us consider the case of collisionless particles, too.
The Poisson equation for the gravitational potential in the case of 
spherical symmetry is 
\begin{equation}
\label{eqpot_sph}
\displaystyle{\frac 1{r^2} \frac d{dr} \left(r^2 \frac{d\Psi}{dr}\right)
	= -4 \pi G \rho}.
\end{equation}
The equilibrium velocity distribution of the particles is a Maxwellian
distribution given by
distribution is
\begin{equation}
f(v) = \displaystyle{\frac{\rho_0}{(2\pi \sigma^2)^{3/2}}
	\exp\left(\frac{\Psi - v^2/2}{\sigma^2}\right)},
\end{equation}
where $\rho_0$ is the central density if we take $\Psi(r=0)=0$, and $\sigma$ is
the one dimensional velocity dispersion.
After integrating over velocity we obtain
\begin{equation}
\rho =\int f(v)d\,{\bf v}= \rho_0 e^{\Psi/\sigma^2}.
\end{equation}
Therefore
\begin{equation}
\Psi = \sigma^2 \ln\left(\frac{\rho}{\rho_0}\right)
\end{equation}
which we substitute in equation (\ref{eqpot_sph}), obtaining
\begin{equation}
\label{rhosig_sph}
\displaystyle{\frac d{dr}\left(r^2 \frac {d(\ln\rho)}{dr}\right)
	= -\frac{4\pi}{\sigma^2} G\rho r^2}.
\end{equation}
By calculating the mean squared velocity we obtain
\begin{equation}
 \langle v^2\rangle=3\sigma^2.
\end{equation}
The equivalent temperature can be calculated from
\begin{equation}
\displaystyle{\frac{\langle v^2\rangle}{2}=\frac 32\frac{k_B T}{m}}
\end{equation}
obtaining for $\sigma$:
\begin{equation}
\label{sigma_sph}
\displaystyle{\sigma^2=\frac{k_BT}m}.
\end{equation}
A comparison of equation (\ref{rhosig_sph}) with equation (\ref{eqrho_sph}) 
using equation (\ref{sigma_sph}) shows they are identical.
Hence, the structure of an isothermal self-gravitating fluid sphere is 
identical with the structure of a collisionless 3D spherical system in 
equilibrium.

To make equation (\ref{rhosig_sph}) nondimensional, we introduce new 
variables
\begin{equation}
\label{vars}
\tilde{\rho} = \displaystyle{\frac{\rho}{\rho_0}}, \qquad
\zeta = \displaystyle{\frac{r}{r_0}},
\end{equation}
where $\rho_0$ is the central density, and we choose
\begin{equation}
\label{r_0} 
r_0=\sigma/(4\pi G\rho_0)^{1/2}.
\end{equation}
 In these variables equation (\ref{rhosig_sph}) becomes
\begin{equation}
\label{nondim_sph}
\frac d{d\zeta}
	\left(\zeta^2 \frac{d(\ln\tilde{\rho})}{d\zeta}\right)
	= -\tilde{\rho} \zeta^2.
\end{equation}
We must solve equation (\ref{nondim_sph})  with boundary conditions
\begin{equation}
\label{init_cond}
\tilde{\rho}(0)=1, \qquad \displaystyle{\frac{d\tilde{\rho}}{d\zeta}(0)=0}.
\end{equation}

A simple expression which satisfies equation (\ref{nondim_sph}) exactly, but
violates the boundary conditions in equations 
(\ref{init_cond}), often used in models due 
to its simplicity, is that of a {\it singular} isothermal sphere, for which
\begin{equation}
\label{rho_sing}
\rho(r)=\displaystyle{\frac{\sigma^2}{2\pi Gr^2}},
\end{equation}
and
\begin{equation}
\label{sigma_sing}
\sigma^2=\frac 12\frac{GM(r)}{r}={\rm constant}.
\end{equation}
However, any solution of both equations (\ref{nondim_sph}) and  
(\ref{init_cond}) must have a core (i.e. 
$r_0\neq 0$), and, while equations (\ref{rho_sing}) and (\ref{sigma_sing})
correspond to the asymptotic limit of the solutions of equations 
(\ref{nondim_sph}) and  (\ref{init_cond}) as $r_0\rightarrow 0$, they
are not good approximations in general. We will compare our results for the 
nonsingular case with that of the singular isothermal sphere in what follows.

The isothermal sphere has infinite mass
($M\propto r$ for large radii). Hence, in order to describe some realistic
finite structure in terms of this model, we must truncate the sphere at some 
radius, $r_t$. The total mass $M_0$ of the isothermal sphere is then
\begin{equation}
M_0=M(r_t)=\int^{r_t}_0 4\pi\rho(r) r^2d\,r=4\pi\rho_0r_0^3 \tilde{M}(\zeta_t)
\end{equation}
where $\zeta_t=r_t/r_0$ and $\tilde{M}$ is the dimensionless mass:
\begin{equation}
\label{mass_nond}
\tilde{M}(\zeta_t)\equiv
	\displaystyle{\frac {M(r_t)}{ 4 \pi r_0^3\rho_0}
	=\int^{\zeta_t}_0\tilde{\rho}(\zeta)\zeta^2 d\zeta}.
\end{equation}
Using equation (\ref{r_0}), we obtain
\begin{equation}
M(r_t)=\frac {\sigma^3}{G(4\pi G \rho_0)^{1/2}}\tilde{M}_t,
\end{equation}
where $\tilde{M}_t\equiv \tilde{M}(\zeta_t)$.
Solving for $\sigma^2$, we obtain
\begin{equation}
\sigma^2=G(4\pi\rho_0)^{1/3}\left(\frac {M_0}{\tilde{M}_t}\right)^{2/3},
\end{equation}
while equation (\ref{sigma_sph}) yields the gas temperature:
\begin{equation}
\label{T-rho}
T = \frac{Gm}{k_B}(4\pi\rho_0)^{1/3}
               \left(\frac {M_0}{\tilde{M}_t}\right)^{2/3}.
\end{equation}

The existence of a truncation radius leads to the necessity of an external 
pressure to keep the system in equilibrium. This requires a significant 
correction to the form of the virial theorem stated in the previous section
 and, accordingly, to all of the other results obtained there.

\section{The Virial Theorem for truncated isothermal spheres and its 
Consequences}
\label{virial}

\begin{figure}
\centerline{\psfig{figure=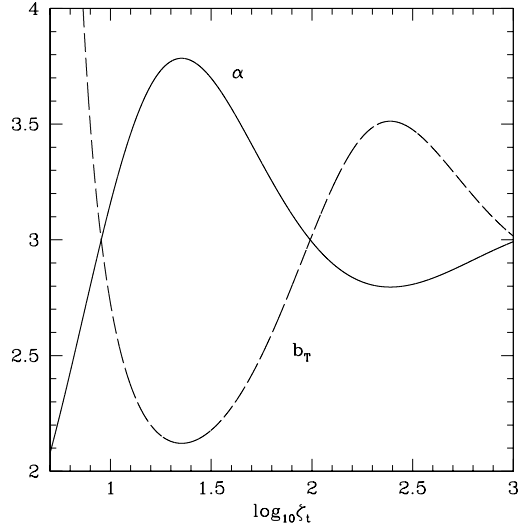,height=3in,width=3in}}
\caption{The ratio of the mean pressure to the boundary pressure
 $\alpha=\bar{p}/p_t$ for
an isothermal sphere truncated at radius $r_t$ (solid curve), is plotted 
against the dimensionless truncation radius $\zeta_t=r_t/r_0$. The ratio of 
the temperature of the isothermal sphere to that of a uniform sphere with 
the same mass and total energy according to the SUS approximation,
$b_T=T/T_{SUS}$, is plotted as well (dashed curve).}
\label{alpha}
\end{figure}

Let us consider the general isothermal sphere density profile $\rho(r)$,
obtained in the previous section. From the ideal gas law, the pressure 
inside as a function of the radius is
\begin{equation}
p(r)=\frac{k_BT}m\rho(r)=\sigma^2 \rho(r),
\end{equation}
and at the outer edge
\begin{equation}
\label{p_t}
p_t= p(r_t)=\sigma^2 \rho(r_t).
\end{equation}
The mean pressure inside the sphere is
\begin{equation}
\bar{p}=\displaystyle{\frac{\int pd\,V}{\int d\,V}}
     =\displaystyle{\frac{3\int^{\zeta_t}_0\tilde{\rho}(\zeta)\zeta^2d\,\zeta}
		{\zeta_t^3\tilde{\rho}(\zeta_t)}p_t}
	=\displaystyle{\frac{3\tilde{M}(\zeta_t)}
		{\zeta_t^3\tilde{\rho}(\zeta_t)}p_t\equiv\alpha(\zeta_t)p_t}
\end{equation}
where $\zeta_t\equiv r_t/r_0$. A plot of $\alpha(\zeta_t)$ is shown on
Figure~\ref{alpha}. The limit $\zeta_t=\infty$, $\alpha=3$ is the case of  
a singular isothermal sphere, which will be considered in more detail below.

The virial theorem for a static sphere in the presence of a surface pressure 
reads
\begin{equation}
\label{vir_th}
0=2K+W+S_p,
\end{equation}
where $W$ is the gravitational potential energy, K is just $U_{th}$, the 
thermal energy, and $S_p$ is the surface pressure term.
The thermal energy for a gas with ratio of specific heats $\gamma=5/3$ is 
given by
\begin{equation}
U_{th}=\frac 32\int pd\,V=\frac 32\alpha(\zeta_t)p_tV,
\end{equation}
where $V$ is the total volume. The surface term is equal to
\begin{equation}
S_p=-\int p{\bf r}\cdot d\,{\bf S}=-3Vp_t.
\end{equation}
Hence the potential energy is
\begin{equation}
W=-2U_{th}-S_p=-2\frac{\alpha-1}\alpha U_{th}
\end{equation}
and the total energy is
\begin{equation}
\label{total_en}
E=\frac{2-\alpha}{\alpha}U_{th}.
\end{equation}
In order to satisfy the virial theorem, the condition that $U_{th}>0$ 
(and $E<0$) requires $\alpha>2$, while Figure~\ref{alpha} indicates that 
$\alpha<3.7846$ in 
general. The virial temperature of the isothermal sphere is
\begin{equation}
\label{temper}
T=\frac {2\alpha}{5(\alpha-2)} \frac{G{M_0}m}{k_Br_m},
\end{equation}
which we shall henceforth refer to as $T_{TIS}$\footnote{Henceforth, the 
notation ``TIS'' shall refer to a solution of the isothermal Lane-Emden 
equation with nonsingular boundary conditions of equation (\ref{init_cond})
at the origin}.
Since $\alpha/(\alpha-2)>1$ for any $\alpha$, the 
temperature $T_{TIS}$ is always higher than $T_{SUS}$, the standard value for
a uniform sphere shown in equation (\ref{temp}). If we write
$T_{TIS}=b_TT_{SUS}$, we find $b_T>2.13$. This temperature correction factor is
also plotted in Figure \ref{alpha}.

The virial radius $r_{\rm vir}$ in this case is just the size of the truncated
isothermal sphere, or $r_{\rm vir}=r_t$. For comparisons with the results of 
dynamical calculations of the formation of such an equilibrium object, we 
should interpret $r_{\rm vir}$ (and $r_t$) as the radius inside which
hydrostatic equilibrium holds \cite{Cole_Lacey}. Care should be taken not to 
confuse this definition of $r_{\rm vir}$ with the one  that is most often used 
previously when identifying collapsed haloes in simulations, that of  the 
radius inside which the mean overdensity is $18\pi^2\approx 178$, as predicted
 by the top-hat solution if $r_{\rm vir}=r_m/2$. We will 
demonstrate below that our solution for $r_{\rm vir}=r_t$ is not far from this 
value, even though the structure of the TIS is very different from that of the
SUS approximation. This is consistent with results obtained using numerical 
simulations \cite{Cole_Lacey}. 
 
For some purposes, the approximation of a truncated, {\it singular}
 isothermal sphere (``SIS'')
has been used, instead of the exact solution, since it is algebraically 
simpler and the results can be obtained analytically, even though this 
approximation cannot be correct at the origin. For comparison with the
nonsingular TIS results and as further illustration of the importance
of the correction for finite boundary pressure discussed above, we calculate 
the virial temperature for such truncated, but {\it singular} isothermal 
spheres as follows. Using equation (\ref{rho_sing}) and the ideal gas law, 
the pressure at radius $r$ is given by
\begin{equation}
p(r)=\displaystyle{p_t\left(\frac{r}{r_t}\right)^{-2}},
\end{equation}
where the boundary pressure at $r=r_t$ is given by
\begin{equation}  
\label{p_t_sing}
p_t=\frac{k_BT}m\rho_t=\sigma^2\rho_t=\frac{\sigma^4}{2\pi Gr_t^2}.
\end{equation}
The average pressure inside the truncated isothermal sphere is then just
\begin{equation}
\label{p_ave}
\bar{p}=3p_t,\qquad 
\end{equation}
(i.e. $\alpha=3$). The thermal energy in this case is given by
\begin{equation}
\label{Uth}
U_{th}=\frac 32\int pd\,V =\frac92 p_tV=\frac{3\sigma^4r_t}G
		=\frac 34 \frac{GM_0^2}{r_t},
\end{equation}
while the surface pressure term in the virial theorem equation is given by
\begin{equation}
\label{Sp}
S_p=-\int p{\bf r}\cdot d\,{\bf S}=-4\pi r_t^3p_t=-\frac{2\sigma^4r_t}G
	=-\frac 12 \frac{GM_0^2}{r_t}.	
\end{equation} 
According to the virial theorem, the potential energy is, therefore, given by
\begin{equation}
W=-2U_{th}-S_p=-\frac{4\sigma^4r_t}G=-\frac 43 U_{th}=-\frac{GM_0^2}{r_t},
\end{equation}
and the total energy becomes
\begin{equation}
E=U_{th}+W=-\frac {U_{th}}3=-\frac 14 \frac{GM_0^2}{r_t}.
\end{equation}
The SIS virial temperature is calculated as above for the nonsingular case
according to equation (\ref{temper}), except that $\alpha=3$ for the singular 
case, which yields
\begin{equation}
\label{temper_sing}
T=T_{SIS}=\frac 65 \frac{G{M_0}m}{k_Br_m},
\end{equation}
This virial temperature is a factor of $b_T=\alpha/(\alpha-2)=3$ times larger
than $T_{SUS}$ in equation (\ref{temp}) given by the SUS approximation. 
Equations (\ref{Uth}) and (\ref{temper_sing}) then yield the truncation 
radius for this SUS case, $r_t=(5/12)r_m$. Hence, in the limiting case of a 
{\it singular}, truncated isothermal sphere, the correction to the virial 
temperature resulting from the finite boundary pressure term is of 
considerable importance, while the difference between the actual radius of the
isothermal sphere and the virial radius of the SUS approximation (i.e.
$r_{vir,SUS}=r_m/2$) is more modest. In what follows, we shall derive the 
corrections for the actual, {\it nonsingular} truncated isothermal sphere 
which results from the top-hat collapse. 
\section{Choosing a Unique Profile: The minimum-energy solution}
\label{min_en}
\subsection{The solution}
As shown above, the truncated isothermal sphere solutions form a one-parameter
family, described by $\zeta_t=r_t/r_0$ -- the truncation radius in units of 
the core radius. Specifying $\zeta_t$, the total mass, and total energy 
completely determines the solution. Alternatively, we can specify 
the mass, the total energy and the applied
external pressure $p_t$. Using these parameters, we can implicitly express the 
truncation radius as follows.
The dimensionless density at $\zeta_t$ is
\begin{equation}
\label{rho_t}
\tilde{\rho}(\zeta_t)=\frac{\rho_t}{\rho_0}=\frac{p_t}{\sigma^2\rho_0}
	=\frac{mp_t}{k_B T\rho_0}.
\end{equation}	
From equations (\ref{r_0}) and (\ref{T-rho}) we obtain
\begin{equation}
\begin{array}{ll}
\label{r_0_rho_0}
r_0&=\displaystyle{\frac{M_0}{\tilde{M}_t}\frac{ mG}{k_B T}},\\
\rho_0&=\displaystyle{\frac 1{4\pi}\left(\frac{\tilde{M}_t}{M_0}\right)^2
	\left(\frac{k_B T}{mG}\right)^3},
\end{array}
\end{equation}
and, by substituting this $\rho_0$ in equation (\ref{rho_t}), we obtain
\begin{equation}
\label{rho_til}
\tilde{\rho}(\zeta_t)=\displaystyle{4\pi\left(\frac{m}{k_BT}\right)^4p_t
	\left(\frac{M_0}{\tilde{M}_t}\right)^2}G^3.
\end{equation}
\begin{figure}
\centerline{\psfig{figure=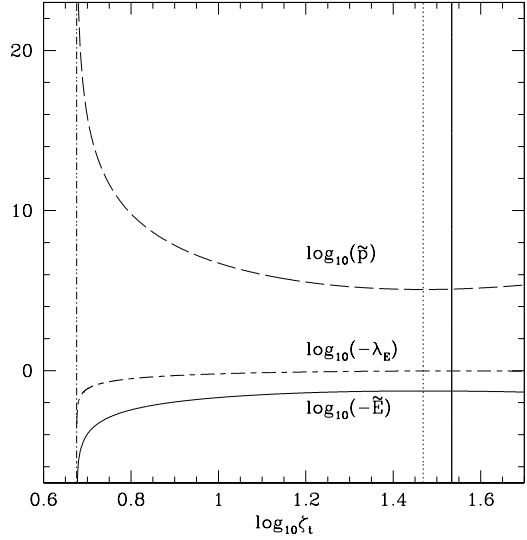,height=3in,width=3in}}
\caption{Logarithms of the dimensionless energy $\tilde{E}(\zeta_t)$ 
(solid curve), external pressure $\tilde{p}(\zeta_t)$ (long-dashed curve), 
and radius $|\lambda_E(\zeta_t)|$ (long-dashed -- short-dashed curve)
versus dimensionless 
truncation radius $\zeta_t$ for a truncated isothermal sphere. The vertical
dot-dashed line indicates the value $\zeta_t=4.738$ below which $E>0$ and a TIS
cannot exist. The vertical dotted line indicates the value $\zeta_t=29.4$ at 
which the energy is minimal. The vertical thick line indicates the boundary of 
stability for the TIS solution ($\zeta_t=34.2$, see \S~\ref{stability}). TIS 
solutions are unstable to the right of this line.}
\label{energy_plot}
\end{figure}
Using equation (\ref{total_en}), the total energy can be written as
\begin{equation}
\label{energy0}
E
	=\frac{2-\alpha}\alpha \frac{3k_B T}{2m}M_0
\end{equation}
and, together with equation (\ref{rho_til}) , we finally obtain
\begin{equation}
\label{energy}
E=\displaystyle{\frac32\left(4\pi G^3 p_t M_0^6\right)^{1/4}
		\tilde{E}(\zeta_t)},
\end{equation}
where	
\begin{equation}
\label{nondim_en}
\displaystyle{\tilde{E}(\zeta_t)=\frac{2-\alpha(\zeta_t)}{\alpha(\zeta_t)
	    \left[\tilde{\rho}(\zeta_t){\tilde{M}^2(\zeta_t)}\right]^{1/4}}}.
\end{equation}
Similarly, we can define a dimensionless external pressure $\tilde{p}$:
\begin{equation}
\label{nondim_press}
\tilde{p}(\zeta_t)\equiv [\tilde{E}(\zeta_t)]^{-4}
\end{equation}
where
\begin{equation}
p_t=\displaystyle{\frac{(2E/3)^4}{4\pi G^3M_0^6}\tilde{p}(\zeta_t)}
\end{equation}
We have plotted the dependence of the dimensionless total energy $\tilde{E}$
and the dimensionless external pressure $\tilde{p}$ on $\zeta_t$
in Figure \ref{energy_plot}. 
In order to indicate the 
dependence of the size of the sphere on $\zeta_t$, we nondimensionalize the 
radius $r_t$, according to
\begin{equation}
\label{lambda_E}
-\lambda_E\equiv\displaystyle{-\frac{r_tE}{GM_0^2}
	=\frac 32\frac{\alpha-2}{\alpha}\frac{\zeta_t}{\tilde{M}_t}}
\end{equation}
and plot $-\lambda_E$ against $\zeta_t$ in Figure~\ref{energy_plot}, as well.
This definition of $\lambda_E$ also corresponds to the familiar dimensionless energy
parameter used in discussions of the stability of isothermal spheres (see 
\S~\ref{stability} below).

For a given mass $M_0$ and external pressure $p_t$, the solution is specified
uniquely only if we can uniquely identify a special value of $\zeta_t$, or
equivalently, of $E$.
Apparently, for any truncated isothermal sphere of mass $M_0$ which is confined
by a given external pressure $p_t$, there is a unique value of $\zeta_t$ which
minimizes the total energy $E$. We shall make the reasonable anzatz that this 
minimum-energy solution is the unique TIS solution preferred in nature as the 
outcome of the virialization of the sphere in the presence of a fixed external 
pressure. We offer evidence to support this anzatz in \S~\ref{Bert} for the
cosmological top-hat problem. It is possible that this is a general result for any
TIS formed by relaxation in the presence of a fixed external pressure, but we are 
only concerned here with spheres that evolve from cosmological initial conditions. 

The minimum value of $E$ as a function of $\zeta_t$ for a given $p_t$ is found 
by minimizing the dimensionless energy $\tilde{E}(\zeta_t)$ in equation 
(\ref{nondim_en}). This occurs for $\zeta_t=29.4$, for which 
$\tilde{E}(\zeta_t)=-0.2816$, $\alpha(\zeta_t)=3.73$, $\tilde{M}_t=61.485$, and
$\tilde{\rho}(\zeta_t)=1.946\times10^{-3}$.
For a given mass $M_0$ and boundary pressure $p_t$, the full details of the 
TIS solution are described in terms of these quantities as follows. 
The total energy $E$ is calculated from equation (\ref{energy}), resulting in
\begin{equation}
E=\displaystyle{\frac32\left(\frac{2-\alpha}{\alpha}\right)
	\left(\frac{4\pi}{\tilde{\rho}_t\tilde{M}_t^2}\right)^{1/4}
		G^{3/4} p_t^{1/4}M_0^{3/2}}
	=-0.795 \,G^{3/4} p_t^{1/4}M_0^{3/2}.
\end{equation}
The temperature $T$ is obtained from equation (\ref{energy0}) and 
(\ref{energy}), according to 
\begin{equation}
\label{T_fin}
T = \left(\frac{4\pi}{\tilde{\rho}_t\tilde{M}_t^2}\right)^{1/4}
	\frac m{k_B}G^{3/4} p_t^{1/4}M_0^{1/2}
		= 14.4\,\frac m{k_B}G^{3/4} p_t^{1/4}M_0^{1/2}.
\end{equation}
Finally, the core radius 
$r_0$ and the central density $\rho_0$ are calculated from equations
(\ref{r_0_rho_0}) and (\ref{T_fin}):
\begin{equation}
\label{r_0_final}
r_0 = \left(\frac{\tilde{\rho}_t}{4\pi\tilde{M}_t^2}\right)^{1/4}
	G^{1/4} p_t^{-1/4}M_0^{1/2}
	= 1.42\times10^{-2}\,G^{1/4} p_t^{-1/4}M_0^{1/2},
\end{equation}
and
\begin{equation}
\label{rho_0_final}
\rho_0
=\left(\frac{\tilde{M}_t^2}{4\pi\tilde{\rho}_t^3}\right)^{1/4}
	G^{-3/4} p_t^{3/4}M_0^{-1/2}
		=449.5\,G^{-3/4} p_t^{3/4}M_0^{-1/2}.
\end{equation}
The size of the TIS is then just 
\begin{equation}
r_t=\zeta_t r_0=29.4\, r_0
\end{equation}
with $r_0$ given by equation (\ref{r_0_final}). The dimensionless radius 
for this value of $\zeta_t$ corresponds then to $\lambda_E=-0.3326$.

Apparently, the core of the truncated isothermal sphere is very small compared
to its overall size. Nevertheless, the profile is quite different from that of
a singular isothermal sphere, in that it has a higher average pressure, lower 
temperature and a core.
%
%
\subsection{Stability}
\label{stability}
The problem of the stability of truncated isothermal spheres has been 
discussed by several authors (e.g. Antonov 1962, Lynden-Bell and Wood 1968, 
Katz 1978, Binney and Tremaine 1987,
Padmanabhan 1989, 1990). According to this literature, our unique minimum 
energy 
solution above is stable, as follows. The case at hand is that of a sphere 
bounded
by a thermally insulating wall, distinct from the case of a sphere in contact 
with
a heat bath. The family of allowed TIS solutions of the Lane-Emden equation 
can be
parameterized by one variable, $\zeta_t$ or, equivalently, by the combination 
of 
parameters given by the ratio of densities measured at $r_0$ and $r_t$,
${\cal R} = \rho_0/\rho_t$ and the dimensionless energy 
$\lambda_E\equiv r_t E/GM^2$.
For gravitationally bound spheres ($E<0$), there are no equilibrium solutions
 for
$\lambda_E<-0.335$. Only for 
$\lambda_E>-0.335$ are isothermal spheres possible. For large enough density 
contrast between the center and edge of the sphere, however, equilibria exist 
but 
are unstable. In particular, as first discovered by Antonov 
\shortcite{Antonov} 
and later discussed in detail in Lynden-Bell and Wood \shortcite{LBW} -- who 
called this ``the gravothermal catastrophe'' -- isothermal spheres with 
$\lambda_E>-0.335$ and ${\cal R} > 708.61$ (corresponding to $\zeta_t=34.2$) 
can exist as solutions of the Lane-Emden equation but are unstable. Such 
spheres are
metastable configurations, for which entropy extrema exist but are not  
local maxima. Only in the intermediate regime, where $\lambda_E>-0.335$ and
${\cal R} < 708.61$, can isothermal spheres both exist and be stable. In this 
regime,
the TIS can be a state of local maximum of the entropy. We note that our 
unique,
minimum-energy solution for the TIS in the presence of a fixed boundary 
pressure 
$p_t$, described above, satisfies the stability requirement for avoiding the 
gravothermal catastrophe, since $\zeta_t=29.40$, 
${\cal R} = [\tilde{\rho}(\zeta_t)]^{-1} = 514$, and $\lambda_E = -0.3326$.

Finally, we note that in the TIS solution derived here, the gas temperature is
spatially {\it uniform} but the equation of state of this gas is {\it not} an
isothermal one (i.e. changes of density for a given fluid element
 do not leave the temperature unchanged).
In the latter case, the stability is analyzed by considering isothermal spheres
in contact with a heat bath, rather than the thermally insulated spheres 
described above. For a gas with an isothermal equation of state, isothermal 
spheres are found to be unstable for density contrasts below that
of the gravothermal catastrophe limit. This analysis does not apply to our
minimum-energy solution above, however.
 Hence, we do not need to
consider the classical results of Ebert (1955, 1957), Bonnor 
\shortcite{Bonnor} and McCrea \shortcite{McCrea} which deal with the problem 
of gas spheres bounded by a given surface pressure in thermal equilibrium with
an external heat bath, which differs from the present case.

\section{The truncated isothermal sphere formed by top-hat collapse}
\label{TIS}
 
\begin{figure}
\centerline{\psfig{figure=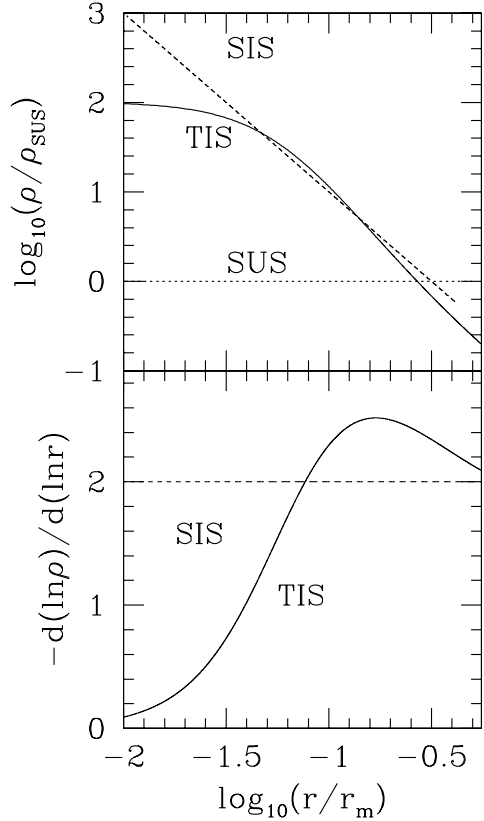,height=5in,width=5in}}
\caption{(upper panel) The postcollapse equilibrium density profile
of a spherical top-hat density perturbation 
versus the radius in units of the radius of the tophat at maximum expansion
for the minimum-energy TIS (solid line), the singular isothermal sphere 
(SIS, dashed line) and the uniform sphere approximation 
(SUS, dotted line). All densities are plotted in units of 
$18\pi^2\rho(t_{\rm coll})$, the postcollapse density of the SUS approximation. 
(lower panel) The logarithmic slope of our minimum-energy TIS profile versus 
radius as compared to the constant slope of $-2$ for the SIS profile. }
\label{dens} 
\end{figure}

The unique TIS solution found in \S~\ref{min_en} by minimizing the total 
energy of the sphere for a given boundary pressure $p_t$  can be applied to 
the top-hat problem, as follows. The virialized object which results from the
collapse of a given top-hat density perturbation must have the mass of the
top-hat and the same total energy as the top-hat before it collapsed and
virialized. Fixing the mass $M_0$ and the energy $E$ of the TIS which
results from the collapse of the top-hat amounts to fixing the boundary
pressure $p_t$ as that value which makes the TIS with this total energy $E$
correspond to the minimum-energy solution described above. Hence, the TIS 
which results from top-hat collapse must have $\zeta_t=29.4$ and $\alpha=3.73$,
as found in \S~\ref{min_en}. This means that the virial temperature $T_{TIS}$ 
is larger than that found by the SUS approach, $T_{SUS}$, by the factor 
$b_T=\alpha/(\alpha-2)=2.16$.
A plot of $\rho/\rho_{SUS}$ as a function of $r/r_t$, demonstrating the
significant difference between the two density profiles is shown on 
Figure~\ref{dens}.

This conservation of the energy $E$ of the top-hat before and after its 
collapse and virialization assumes that there is no extra $pdV$-work that 
needs to be taken
into account due to the presence of the external boundary pressure $p_t$ which
would otherwise alter the final total energy $E$ compared with its initial 
value 
before collapse. This is appropriate for the case at hand, since the collapse 
prior
to the epoch of virialization is that of a cold, pressure-free gas which 
collapses
supersonically. As in the well-known, self-similar, spherical infall solution
 of
Bertschinger (1985), the original energy is converted from potential energy at
maximum expansion to a mixture of infall kinetic energy and potential energy 
during
infall, with a negligible share of the energy going into compressional 
heating. In
that solution, the infall is halted by a strong shock. At this shock, the 
kinetic
energy of infall is converted primarily into the thermal energy of the 
shock-compressed gas, with only a small portion remaining as kinetic energy of
subsonic, postshock infall. By analogy with this infall solution, therefore, we
identify the boundary pressure $p_t$ in the case of our TIS solution, not as a
 fixed
external pressure which acts on the top-hat boundary throughout its collapse 
and 
virialization, but rather as something like the instantaneous post-shock 
pressure
in the infall solution. As such, it has a physical origin in the conversion of
 the 
original energy of the collapsing top-hat, itself, from potential energy at 
maximum expansion into kinetic energy of infall during collapse and, finally,
into thermal energy of the post-shock gas, always conserving the original 
energy $E$
of the top-hat.  
 
\begin{figure}
\centerline{\psfig{figure=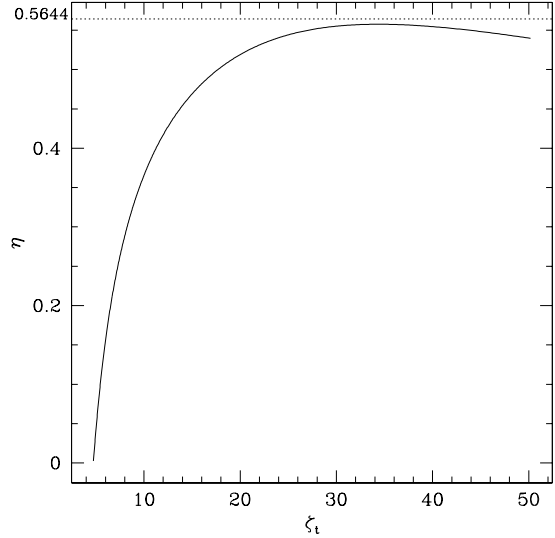,height=3in,width=3in}}
\caption{The collapse factor relating the radius of the initial top-hat at its 
point of maximum expansion to its radius after collapse and virialization,
$\eta=r_t/r_m$, as function of the truncation radius $\zeta_t$.
The dotted line indicates the value of $\eta$ given by the self-similar infall
solution (see \S\ref{Bert}).}
\label{eta_fig}
\end{figure}

The final size of the sphere after collapse and virialization is related to 
the 
size of the parent top-hat density fluctuation at the epoch of turnaround by a 
collapse factor $\eta$ defined by
\begin{equation}
\label{eta}
\eta=\frac{r_t}{r_m}.
\end{equation}
Using equations (\ref{eta}) and (\ref{lambda_E}), it is easy to see that
\begin{equation}
\label{eta_zeta}
\eta=-\frac 53\lambda_E=\frac 52\frac{\alpha(\zeta_t)-2}{\alpha(\zeta_t)}
	\frac{\zeta_t}{\tilde{M}(\zeta_t)}
\end{equation}
A plot of $\eta(\zeta_t)$ versus $\zeta_t$ is shown in Figure~\ref{eta_fig}.
For the minimum-energy  solution, using $\zeta=\zeta_t=29.4$ in equation
(\ref{eta_zeta}) yields $\eta=0.554$, i.e. $r_t=0.554\, r_m$. This value of
$\eta$ derived for the minimum-energy TIS solution above is  
similar to the value usually cited for the postcollapse, virialized object
in the SUS approximation, for which $r_{\rm vir}=r_m/2$, corresponding to 
$\eta_{SUS}=0.5$. It is somewhat larger than the value for the limiting case of
a {\it singular} isothermal sphere (``SIS''), for which $\eta=0.4167$. 

The radius $r_t$ of the TIS solution is encountered by the outer boundary of 
the collapsing top-hat at a time somewhat earlier than the time $t_{\rm coll}$ of 
infinite density in the top-hat solution. We shall
refer to the time at which the radius of the collapsing top-hat equals $r_t$
as $t_{\rm cross}$, to distinguish it from $t_{\rm coll}$. At time $t_{\rm cross}$, the 
top-hat exact solution (\ref{exact}) yields an overdensity of 
$\delta=\delta_{\rm cross}\approx 100$, while the extrapolated linear solution
at this time yields $\delta_L=\delta_{L,cross}\approx 1.52$. If we assume that
the TIS forms instantaneously at $t=t_{\rm cross}$, then its mean density
corresponds to a mean overdensity $\bar{\delta}=\bar{\delta}_{\rm cross}\approx 
100$ when compared with the background density at $z_{\rm cross}$, and corresponds
to a mean overdensity $\bar{\delta}=\bar{\delta}_{\rm coll}\approx 130$ when 
compared with the background density at 
$z_{\rm coll}$. This latter value differs somewhat from the conventional value of 
$18\pi^2\approx 178$, which is found in the SUS approximation. 

In applications of the top-hat model involving the Press-Schechter
approximation, it is customary to identify the characteristic time of formation
of objects of a given mass in terms of $\dc$ in equation (\ref{deltas}). 
Since $\delta_{L,cross}<\dc$, it may be appropriate to replace $\dc$ in such
applications by the value $\delta_{L,cross}\approx 1.52$, implying that the 
number of objects formed at any epoch with a mass greater than some value will
be somewhat higher for the TIS solution than previously assumed in applications
of the Press-Schechter approximation. We note that comparisons of the 
predictions of the Press-Schechter approximation for the mass function of
virialized haloes at any epoch with the results of $N$-body simulations of 
cosmic structure formation from scale-free initial conditions have sometimes
indicated 
better agreement if the standard value of $\dc=1.69$ is replaced by a lower
value of $\dc$ (e.g. Efstathiou and Rees 1988; Carlberg and Couchman 1989;
Klypin  et al. 1995; Bond and Myers 1996; Cen 1998).
Perhaps our TIS result that $\delta_{L,cross}\approx 1.52$ is part of the 
explanation for this.

In order to check our derivation of the unique TIS solution described above and
confirm the validity of the minimum-energy argument on which it is based, we
take a second approach to the same problem, in the following section. In this 
second approach we match the properties of the final TIS equilibrium 
configuration to those of the post-shock gas in the well-known, frequently 
applied, self-similar, cosmological, spherical infall solution \cite{Bert85}.
As we shall see, this approach not only confirms our derivation above, but also
provides a dynamically self-consistent origin for the final equilibrium state
of the TIS.
\section{Application of the Self-Similar Spherical Infall Solution}
\label{Bert}
The description of the postcollapse virial equilibrium of a top-hat density 
perturbation as a TIS with the same total mass and energy as the parent top-hat is
not unique without the use of additional information which establishes the radius,
the temperature, or the boundary pressure. We have argued above that a natural
outcome of the dynamical relaxation of the collapsed top-hat is that unique TIS 
which makes its energy $E$ correspond to a minimum-energy solution at fixed 
boundary pressure. This, in turn, establishes what the value of $p_t$ is for
the TIS which results from the collapse of a given top-hat, and, thereby, fixes
the unique TIS solution. To test this anzatz, we shall now place our top-hat 
density perturbation and its collapse and virialization in a dynamically 
self-consistent context, by comparing with the closely related problem of the
spherical infall which results when a point-mass perturbation is added to a 
matter-dominated Einstein--de Sitter universe with initially uniform density, for
which an exact solution is possible. This latter problem represents the simplest
kind of generalization of a uniform top-hat initial condition which can lead to an
interruption of the collapse of the entire perturbation to infinite density at a 
finite time, due to the radial nonuniformity of the initial conditions. As such,
this problem can be used to provide some insight into how inhomogeneities in a 
collapsing top-hat might halt its collapse prior to its reaching infinite density,
causing it to thermalize the kinetic energy of collapse and relax to an equilibrium
configuration. The spherical symmetry of this point-mass perturbation problem,
moreover, means that one can identify a corresponding top-hat perturbation for
the mass interior to any spherical shell in the point-mass 
perturbation problem, by defining the top-hat which contains the same total mass 
and energy inside the same size sphere at the same initial moment. Hence, a 
quantitative comparison of the solution of this 
problem and a suitably chosen, matching top-hat problem should be possible. We 
describe this in what follows. 

\subsection{The self-similar spherical infall solution}
Consider the self-similar spherical infall solution for the flow which results 
when a point-mass perturbation is introduced at some point in an otherwise 
unperturbed Einstein-de Sitter universe \cite{Bert85}. For a collisional fluid,
this solution involves an accretion shock which occurs at radius 
$r_S=\lambda_S r_{ta}$, where $r_{ta}$  is the radius of that spherical shell
of matter surrounding the initial perturbation which is just turning around at
any epoch $t$ and where $\lambda_S$ is a constant. For an ideal gas with ratio
of specific heats $\gamma=5/3$, the constant $\lambda_S\approx0.34$.  As we 
shall describe in more detail below, it is appropriate for us approximately to
identify the region
of this self-similar infall which is bounded by the accretion shock at radius 
$r_S$ with our postcollapse equilibrium sphere. In order to do this, we must
make the appropriate correspondence between the top-hat perturbation in our
initial conditions and the ``matching'' top-hat in the infall solution which
results when we add the initial point-mass perturbation to the unperturbed gas
in a finite sphere so as to encompass the same mass in the infall solution as
in our initial top-hat. The correct ``matching'' of the two solutions must
ensure that the spherical shell which is just reaching the shock in the infall
solution at some time $t$ was following the same trajectory as the outer
boundary of our top-hat perturbation at all times prior to the time it 
encounters the shock at $r_S$. We outline this ``matching'' solution below. We
argue, further, that this collisional fluid solution should apply equally
well to the case of a TIS in a gas of cold, collisionless dark matter or any 
combination of fluid with $\gamma=5/3$ and cold, collisionless gas, as follows.
Interestingly enough, for the same kind of perturbation in an unperturbed,
purely collisionless, initially cold gas (e.g. cold dark matter), the 
self-similar solutions of Bertschinger (1985) and Fillmore and Goldreich (1984)
yield an inner region of shell crossings, bounded by the outermost caustic
located at $r_{c,1}=\lambda_{c,1} r_{ta}$, where $\lambda_{c,1}\approx0.36$.
As a result, the outer boundaries of the shocked gas and of the region of the 
shell-crossings and caustics in the collisionless dark matter are almost
coincident, and their respective mass distributions are unbiased with respect 
to each other. In that case, we can focus in what follows on the 
Bertschinger \shortcite{Bert85} solution
for a collisional fluid with $\gamma=5/3$ and $\Omega=1$, but bear in mind that
our final result for the TIS applies equally well to a collisionless dark
matter halo or to a combination of the two, as well, as long as $\Omega=1$.

Following Bertschinger \shortcite{Bert85}, we define the dimensionless 
variable $\lambda=r/r_{ta}(t)$, where $r_{ta}$ is the radius of the shell that
is just turning around at the time $t$.
The fluid variables pressure $p$, mass density $\rho$, and mass $m(r)$ 
interior to
radius $r$, are non-dimensionalized using the following definitions of 
dimensionless
pressure $P$, density $D$ and mass $M$ interior to dimensionless radius 
$\lambda=r/r_{ta}$:
\begin{eqnarray}
\label{dimens}
p(r,t)
	=\rho_b\displaystyle{\left(\frac {r_{ta}}t\right)^2}P(\lambda),
		\qquad
\rho(r,t)
	=\rho_bD(\lambda),\qquad
m(r,t)
	=\frac 43\pi\rho_b\,r_{ta}^3M(\lambda),
\end{eqnarray}
where $\rho_b$ is the mean, unperturbed background density at time $t$.
Inserting these definitions into the fluid conservation partial differential 
equations leads to a set of ordinary differential equations. These equations, 
together with strong, adiabatic shock jump boundary conditions, the exact, 
pressure-free infall solution for the flow exterior to the shock and the assumption
that the energy of the gas bounded by the shock is the same as the initial energy of
the same mass when it was outside the shock, describe a self-similar infall solution
with an accretion shock. Bertschinger (1985) presented detailed results of his
numerical solution of these equations both graphically and in tabular form. 
However, since we shall need a finer grid of solution values than is tabulated 
there, we have re-integrated the resulting equations ourselves for the current 
paper, using the stiff equation
solver RADAU5 by E. Hairer and G. Wanner \cite{Hairer}.

The shock occurs at 
fixed $\lambda=\lambda_S\equiv r_S/r_{ta}$ and propagates according to
\begin{equation}
r_S(t)=\lambda_Sr_{ta}(t)\propto t^{8/9}.
\end{equation}
The mass inside the shock is determined by:
\begin{equation}
m(r_S,t)=\frac 43\pi\rho_b r_{ta}^3(t)M(\lambda_S),
\end{equation}
The self-similar infall solution yields, for $\gamma=5/3$:
\begin{equation}
\begin{array}{rr}
 \lambda_S=0.338976, \qquad M(\lambda_S)=3.78759,\qquad
  P(\lambda_S)=9.73563, \qquad D(\lambda_S)=16.5341 	\,.
\end{array}
\end{equation}
%
%
The mean overdensity inside the shock relative to the mean unperturbed 
background density is constant in time, and has a value
\begin{equation}
\label{mean_rho_B}
\bar{\rho}_S=\displaystyle{\frac{m(r_S,t)}{\frac 43 \pi r_S^3\rho_b}
	=\frac{M(\lambda_S)}{\lambda_S^3}=97.24.}
\end{equation}

\begin{figure}
\centerline{\psfig{figure=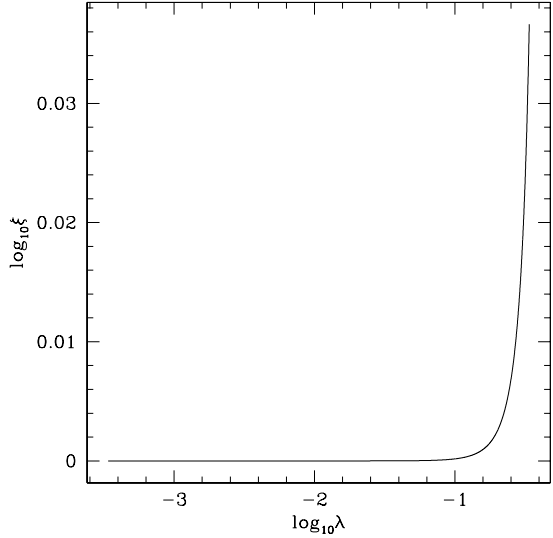,height=3in,width=3in}}
\caption{The ratio of pressure force to gravitational force inside
the accretion shock in the self-similar infall solution. A value $\xi=1$ 
implies hydrostatic equilibrium. The shock is located at 
$\log_{10}\lambda_S=-0.469831$.}
\label{xi}
\end{figure}

Consider the post-shock gas in this self-similar infall solution.
Strictly speaking, this postshock flow is neither isothermal nor in hydrostatic 
equilibrium. As we shall see, however, the variation of
temperature with radius is rather weak, so isothermality is a reasonable 
approximation and 
the post-shock gas is in approximate hydrostatic equilibrium, as well.
We shall demonstrate this as follows.
If the gas is in hydrostatic equilibrium, it must satisfy the equation
\begin{equation}
\frac 1\rho\frac{dp}{dr}=-\nabla\phi.
\end{equation}
Let us define a dimensionless ratio $\xi$,
\begin{equation}
\label{xi1}
\xi\equiv \frac 1{\rho}\frac{dp}{dr}\big/(-\nabla\phi),
\end{equation}
which measures the degree to which hydrostatic equilibrium is obtained at any 
radius $r$, so that $\xi=1$ if equilibrium occurs.
In order to demonstrate how close the exact solution is to hydrostatic 
equilibrium 
inside the shock, we rewrite $\xi$ in terms of the dimensionless fluid 
variables,
according to
\begin{equation}
\label{xi_def}
\xi(\lambda) = \frac {9}2
	\left[\frac{\lambda^2 P'(\lambda)}{M(\lambda)D(\lambda)}\right].
\end{equation}
We use the exact numerical solution to plot $\xi(\lambda)$ in Figure~\ref{xi}. 
Apparently, the postshock gas is always close to hydrostatic, with a maximum 
departure of $\xi$ from unity of 9 per cent just behind the shock, where the gas still
has a small infall velocity and the pressure is a little lower than the value
required to balance gravity.
  
The approximation
of equilibrium becomes perfect as we approach the origin.
We can demonstrate this analytically,
using the simple asymptotic scalings of the fluid variables with 
distance there. In terms of $\lambda$, the solution obeys the following 
scalings at small $\lambda$:
\begin{equation}
\label{scalings}
P=\tilde{P}(\lambda)\lambda^{-5/2},\,\,\,\,
D=\tilde{D}(\lambda)\lambda^{-9/4},\,\,\,\,
M=\tilde{M}(\lambda)\lambda^{3/4},
\end{equation}
where $\tilde{P}$, $\tilde{D}$ and $\tilde{M}$ are only weakly dependent on 
$\lambda$ for small $\lambda$, $\tilde{P}(0)=2.40676$, 
$\tilde{D}(0)=2.60173$, and $\tilde{M}(0)=10.4069$.
Inserting equation (\ref{scalings}) into equation (\ref{xi_def}) for 
$\xi(\lambda)$ shows that $\xi$ is independent of $\lambda$ as $\lambda$ 
approaches 0, with $\xi(0)=1$.   

The postshock gas is also not far from isothermal, despite the strong increase of 
pressure and density towards the centre. The volume-averaged temperature interior 
to the shock radius is related to the immediate postshock temperature $T_S$ 
according to
\begin{equation}
\label{T_V}
\displaystyle{\frac{\langle T\rangle_V}{T_S}=3\lambda_S^{-3}\int^{\lambda_S}_0
	\frac{P(\lambda)/P(\lambda_S)}{D(\lambda)/D(\lambda_S)}\lambda^2d\lambda}.
\end{equation}
According to the scaling laws in equation (\ref{scalings}), valid at small
radii and approximately correct throughout much of the postshock region, temperature
varies with radius only as $r^{-1/4}$. In terms of the scaling laws,
\begin{equation}
\label{T_Vsca}
\displaystyle{\frac{\langle T\rangle_V}{T_S}\approx \frac{12}{11}
	\lambda_S^{-1/4}
	\frac{\tilde{P}(0)/P(\lambda_S)}{\tilde{D}(0)/D(\lambda_S)}
		\approx 2.25,}
\end{equation}
 while the exact solution yields
\begin{equation}
\label{T_Vnum}
\displaystyle{\frac{\langle T\rangle_V}{T_S}=1.386,}
\end{equation}
which is remarkably close to unity.

In short, the self-similar infall solution indicates that the dynamically 
self-consistent evolution of an uncompensated, spherically-symmetric, positive, 
cosmological density perturbation in a cold gas leads to continuous infall and
the creation of a continuous
sequence of spheres of collapsed gas which are, at any instant, in a state which
is close to that of an isothermal sphere in hydrostatic equilibrium, bounded by
the accretion shock. In what follows, we identify the particular self-similar,
shock-bounded sphere in this solution which instantaneously corresponds to the 
postcollapse,
virialized sphere which forms as a result of the equivalent top-hat perturbation.
\subsection{Matching the post-shock gas in the infall solution with
the postcollapse TIS of the top-hat}

In what follows, we attempt to find the TIS solution for the 
postcollapse virial equilibrium of a top-hat density perturbation which
most closely matches the characteristics of the accretion-shock-bounded sphere in 
the self-similar spherical infall solution when the mass and energy inside the 
latter sphere equal the mass and energy of the top-hat. We 
require that the trajectory of the top-hat outer boundary be the same as that of the
immediate postshock fluid element in the infall solution during the preshock 
infall of the latter.
This fixes the radius of the shock and the time at which the shock encloses a mass
equal to that of the corresponding top-hat. By comparing this shock radius with the 
radius of the immediate postshock fluid element at its point of maximum expansion,
we shall derive the effective collapse factor for the corresponding top-hat from
its maximum expansion to its postcollapse virial equilibrium state, by identifying
the shock radius $r_S$ as the radius $r_t$ of the matching TIS. As we shall see,
this infall shock radius in almost precisely equal to the radius derived previously
for the minimum-energy TIS solution, although it is necessary to make a tiny
downward adjustment of the value of $r_t$ relative to $r_S$ in order to match a TIS
solution with the same total mass and energy. In particular, an isothermal sphere 
with exactly the same mass and energy as this shock-bounded sphere in the infall
solution can only be in hydrostatic equilibrium (i.e. a solution of the Lane-Emden
equation) if its radius is a tiny bit smaller than the shock radius in the infall 
solution. We shall make an approximation, therefore, to determine the actual
radius $r_t$, slightly below $r_S$, of the equilibrium TIS solution which most 
closely matches the shock-bounded sphere in the infall solution, as follows. As we
shall see, the resulting TIS is not far from our TIS solution
derived above from the minimum-energy argument, so we can be confident that our
minimum-energy TIS solution has a sound physical basis. We begin by solving for the
shock radius $r_S$ and time $t_{\rm cross}$ at which the collapsing outer boundary of 
the top-hat reaches the radius of the shock in the infall solution, when the gas 
enclosed by the shock in that solution contains the same mass $M_0$ and energy $E$ 
as the top-hat.

The total mass of the top-hat we consider is:
\begin{equation}
\label{M0}
M_0 = \frac{4\pi}3 r_m^3 (1+\delta_m) \rho_b (t_m)	\,.
\end{equation}
We require that the masses of the top-hat and the post-shock gas in the 
self-similar infall 
solution be equal at the time when the outer radius of the top-hat crosses
 the radius of the shock. This yields
\begin{equation}
\label{M}
M_0 = m(r_S, t_{\rm cross}) = \frac43 \pi \, \rho_b(t_{\rm cross}) \, 
		r_{ta}^3 (t_{\rm cross}) \, M(\lambda_S),
\end{equation}
and, therefore, 
\begin{equation}
\label{r_ta}
r_{ta} (t_{\rm cross}) = \left[ \frac{M_0}{M(\lambda_S)} 
	\frac3{4\pi \rho_b (t_{\rm cross}) } \right]^{1/3},
\end{equation}
where 
\begin{equation}
\label{rhob}
\rho_b(t) = (6\pi G t^{2} )^{-1},\,\,\,\,
\rho_b(t_{\rm cross})=\rho_{b0}(1+z_{\rm cross})^3,
\end{equation}
and $\rho_{b0}=3H_0^2/8\pi G$, the mean background density at present.
Combining equations (\ref{M0}) -- (\ref{rhob}), we can write
\begin{equation}
\label{r_ta_t}
\displaystyle{\left[\frac {r_{ta}(t_{\rm cross})}{t_{\rm cross}}\right]^2}
	=\frac 32\frac{(36\pi)^{1/3}G}{[M(\lambda_S)]^{2/3}}
		M_0^{2/3}(1+z_{\rm cross})\rho_{b0}^{1/3},
\end{equation}
and
\begin{equation}
\label{r_m}
r_m=\displaystyle{\left(\frac 3{4\pi}\right)^{1/3}
	\frac{\lambda_S}{\eta [M(\lambda_S)]^{1/3}}
		M_0^{1/3}(1+z_{\rm cross})^{-1}\rho_{b0}^{-1/3}}.
\end{equation}
The radius of the shock wave at $t_{\rm cross}$ is, using equations (\ref{deltas}) and
(\ref{r_ta}), given by
\begin{equation}
\label{r_S}
\begin{array}{ll}
r_S(t_{\rm cross})&=\displaystyle{\lambda_S r_{ta}(t_{\rm cross})}
	=\displaystyle{\frac{\lambda_S}{[M(\lambda_S)]^{1/3}}(1+\delta_m)^{1/3}
	\left( \frac{t_{\rm cross}}{t_m} \right)^{2/3} r_m}  \\&
	=\displaystyle{\frac{\lambda_S}{[M(\lambda_S)]^{1/3}}
		(1+\delta_m)^{1/3}
	\frac{\delta_{L,cross}}{\delta_{Lm}} r_m}
	=0.36244\delta_{L,cross}r_m,
\end{array}
\end{equation}
where $\delta_{L,cross}$ is the linear density contrast corresponding to the 
time $t_{\rm cross}$.
Expressed in terms of $r_m$, this yields, 
\begin{equation}
\label{r_tS}
r_S(t_{\rm cross})=\eta_S r_m,
\end{equation}
where $\eta_S=0.564393$ \cite{Bert85}. 

Our goal here was to use the shock radius $r_S(t_{\rm cross})$ to establish $r_t$,
the size of our postcollapse equilibrium TIS, by equating $\eta$ in equation 
(\ref{eta}) with $\eta_S$ in equation (\ref{r_tS}).
This value of $\eta_S$ is extremely close to the value of $\eta=0.554$ derived 
above 
for the unique minimum-energy TIS solution for the postcollapse equilibrium of the
top-hat. Such an agreement is remarkable and can hardly be a coincidence. We take
this as a strong confirmation of the validity of that minimum-energy solution.
However, as mentioned earlier, we also note that this tiny difference between
$\eta=0.564$ and $\eta=0.554$ is nevertheless enough to put the isothermal sphere
with mass $M$, energy $E$ and radius $r_t=r_S$ ever so slightly beyond the 
boundary of allowed solutions of the Lane-Emden equations, according to 
Figure~\ref{eta_fig} and equation (\ref{eta_zeta}). In other words, since
$\lambda_E=-(3/5)\eta_S=-0.338636<-0.335$ in this case, such an isothermal 
sphere is just slightly too large to be an equilibrium TIS. In what follows we
shall refine this estimate of the TIS radius $r_t$ so that it
departs from $r_S$ only minimally but nevertheless preserves the physical basis in 
terms of the self-similar infall solution.

\subsubsection{The TIS solution whose size is closest to the shock radius in 
the infall solution}

The significance of the remarkably close agreement  between the radius 
$r_S(t_{\rm cross})$ derived above and the radius $r_t$ of the minimum-energy TIS 
for the same parent top-hat is made clear by the curve of $\eta$ versus 
$\zeta_t$ in
Figure~\ref{eta_fig} which shows that $\eta=0.554$, the value for $\zeta_t=29.4$, 
the
minimum-energy TIS solution value is close to the maximum value of $\eta$ for all
values of $\zeta_t$, while $\eta_S=0.564$ is only just above the curve. 
Since the postshock flow is only approximately isothermal and hydrostatic it is
reasonable to suppose that the final relaxation to equilibrium keeps the total 
energy and mass constant but allows the radius to shrink slightly, just enough to
relax to a stable equilibrium TIS.
The simplest way to refine the value of $r_t$ (or, equivalently, of $\eta$) 
downward from $r_S$ (or $\eta_S$) so as to make an equilibrium TIS solution 
possible, therefore, is just to
take $\eta=\eta_{max}$, the maximum value of $\eta$ versus $\zeta_t$ and, hence, 
the allowed value which is closest to $\eta_S$. This yields the estimate 
$\eta_1=\eta_{max}=0.558$, for which $\zeta_t=\zeta_{t,1}=34.2$ and 
$\alpha=\alpha_1\equiv3.66$. This $\eta_{max}$ solution corresponds exactly to the
unique stable equilibrium TIS for which $\lambda_E=-0.335$ and ${\cal R}=708.61$.
This value of $\eta=\eta_1$ differs from
$\eta_S$ by only $1.3$ per cent. With this choice of $\eta_1$, the TIS solution is
extremely close to that from the minimum-energy solution, with $\eta_1$ differing 
by only $0.7$ per cent, $\alpha$ and $T$ for the two solutions differing by less than 
$2$ per cent for each, and $\zeta_t$ (i.e. the core radius $r_0$) by only 
$17$ per cent.

This agreement between our minimum-energy TIS solution and the infall 
accretion-shock size matching TIS solution confirms the validity of our 
minimum-energy TIS solution. We note, however, that even the very small 
difference found between the two TIS solutions is just enough to place the 
size-matching solution
on the boundary, described above, of gravothermal catastrophe. Apparently,
self-similar spherical infall leads naturally to the formation of a shock-bounded 
sphere which is close to that of an isothermal sphere on the edge of the 
 gravothermal catastrophe regime.

%
%
\section{Discussion and Summary}
\label{results}

\subsection{Summary of the minimum-energy TIS solution}
\label{summary_tis}

We have argued above that a cosmological top-hat density perturbation in an 
Einstein-de Sitter universe which virializes following its collapse can be
described self-consistently by a unique TIS solution which conserves the 
energy of the initial top-hat and is bounded by a pressure which makes that 
energy a minimum. The temperature derived in this way for the sphere in
virial and hydrostatic equilibrium is larger by a factor of approximately
two than the value previously derived by satisfying energy conservation and 
the virial theorem for a postcollapse sphere of uniform density for
which the surface pressure term is neglected. The origin of the boundary
pressure term in the TIS solution can be understood self-consistently by
comparison with the well-known self-similar, spherical infall solution of
Bertschinger\shortcite{Bert85} for the problem of the collapse of matter onto
a point-mass perturbation added to a uniform, cosmologically expanding 
background. In the latter problem, spherical mass shells surrounding the point
mass decelerate relative to the mean cosmic expansion, reach a maximum radius
and collapse back toward the center supersonically, unaffected by pressure 
forces, following a trajectory identical to that of the outer boundary of a 
top-hat perturbation which encompasses the same total mass and energy. Unlike
the collapse of the top-hat, however, the self-similar infall is interrupted
by a strong accretion shock which decelerates the flow and thermalizes the 
kinetic energy of infall, allowing the postshock gas to settle subsonically 
into a state which is approximately hydrostatic and isothermal. By comparing 
the shock-bounded sphere in this self-similar infall solution to the
unique minimum-energy TIS solution which we derived here by a completely 
independent line of reasoning, we have been able to confirm the validity of the
TIS solution while explaining its dynamical origin and the nature of the
boundary pressure. The boundary pressure, in short, is closely related to the 
presence of the accretion shock in the self-similar infall solution. While
the TIS boundary pressure is not identical to the instantaneous postshock 
pressure in the infall solution (it is slightly higher), the TIS boundary 
pressure corresponds to the slightly adjusted postshock pressure required to
transform the quasi-hydrostatic postshock flow into an exact hydrostatic
equilibrium while conserving the total energy inside the sphere. We have also
used the properties of the self-similar infall solution in the case in which
both collisionless dark matter and a collisional fluid are present to argue
that we can apply our TIS solution to either component or both. For the
collisionless component, velocity dispersion takes the place of temperature.
If both components are present, then the density of each component $i$, 
$\rho_i$, is simply proportional to the total, $\rho_i=\rho_{tot}\Omega_i$,
where $\Omega_i$ is the universal density parameter for component $i$.

The TIS solution for the postcollapse equilibrium of a top-hat of mass $M_0$
which would collapse at epoch $z_{\rm coll}$ if its collapse were not interrupted 
by virialization depends upon $M_0$, $z_{\rm coll}$, $h$, the mean total 
background density $\rho_{b,0}$ (where $\rho_{b,0}\propto h^2$), and the 
ratio $\zeta_t$ of truncation radius $r_t$ to core radius $r_0$, according to
\begin{eqnarray}
\label{scaling1}
 T
	&\propto& \displaystyle{\frac\alpha{\alpha-2}}{M_0}^{2/3}(1+z_{\rm coll})
	h^{2/3},\\
\label{scaling2}
 v_c&\propto&\eta^{-1/2}\displaystyle{M_0^{1/3}(1+z_{\rm coll})^{1/2}h^{1/3}},\\
\label{scaling3}
 r_m
	&\propto&  M_0^{1/3}(1+z_{\rm coll})^{-1}h^{-2/3},\\
\label{scaling4}
\rho_0&\propto&\frac{\zeta_t^3}{\eta^3}\frac 1{\tilde{M}_t}
		\rho_{b0}(1+z_{\rm coll})^3,
\end{eqnarray}
where $\alpha$ and $\tilde{M}_t$ depend only on $\zeta_t$ and where $h$ is the 
Hubble constant in units of 100 ${\rm km}$ $\rm{s^{-1}\,Mpc^{-1}}$. 
The radii $r_t=\eta r_m$ and $r_0=r_t/\zeta_t$ have the same scaling with
$M_0$, $1+z_{\rm coll}$, and $h$ as does $r_m$. 

The value we have derived above for $\zeta_t$ for our minimum-energy TIS 
solution is $\zeta_t=29.4$. All of the parameters of this unique TIS solution 
are summarized in Table~\ref{summary}. Using this value and evaluating the 
proportionality constants in the scaling laws (\ref{scaling1}) -- 
(\ref{scaling4}), the scaling laws become the following:
\begin{eqnarray}
r_m &=& \displaystyle{\left(\frac{8G}{9\pi^2}\right)^{1/3}
		M_0^{1/3}(1+z_{\rm coll})^{-1}H_0^{-2/3}}
     = 338.3 \left(\frac M{10^{12}M_{\odot}}\right)^{1/3}(1+z_{\rm coll})^{-1}
		h^{-2/3}\,\, {\rm kpc},\\ 
r_t &=& \eta r_m = 187.4 \left(\frac M{10^{12}M_{\odot}}\right)^{1/3}
	(1+z_{\rm coll})^{-1}h^{-2/3}\,\, {\rm kpc},\\
T &=&\frac{(3\pi)^{2/3}}5\frac{\alpha}{\alpha-2}\frac{m_p}{k_B}G^{2/3}\mu
	M_0^{2/3}(1+z_{\rm coll})H_0^{2/3}  
	= 1.328\times 10^6\mu\left(\frac {M_0}{10^{12}M_{\odot}}\right)^{2/3}
	(1+z_{\rm coll})h^{2/3}\,\, {\rm K},\\
\sigma^2 &=&\frac{k_BT}{m}
	=\frac{(3\pi)^{2/3}}5\frac{\alpha}{\alpha-2}G^{2/3}
	M_0^{2/3}(1+z_{\rm coll})H_0^{2/3}  
	=1.096\times10^4\left(\frac {M_0}{10^{12}M_{\odot}}\right)^{2/3}
	(1+z_{\rm coll})h^{2/3}\,\, {\rm km^2\,s^{-2}},\\
v_c&=&\displaystyle{\left(\frac{GM_0}{r_t}\right)^{1/2}} 
	= 151.4\left(\frac {M_0}{10^{12}M_{\odot}}\right)^{1/3}
	(1+z_{\rm coll})^{1/2}h^{1/3}\,\, {\rm km/s},\\
r_0&=&\frac{r_t}{\zeta_t}
	= 6.37\left(\frac {M_0}{10^{12}M_{\odot}}\right)^{1/3}(1+z_{\rm coll})^{-1}
		h^{-2/3}\,\, {\rm kpc},\\
\rho_0&=& 6\pi^2\displaystyle{\left(\frac{b_T}5\right)^3
		\tilde{M}_t^2(1+z_{\rm coll})^3\rho_{b0}}
	=1.796\times10^4(1+z_{\rm coll})^3\rho_{b0}
	=3.376\times 10^{-25}(1+z_{\rm coll})^3h^2 \,\,{\rm g/cm^3}.
\end{eqnarray}
%
%
\begin{table}
\caption{Summary of the minimum-energy TIS solution}
\label{summary}
\begin{tabular}{@{}ll}
Quantity& Value\\ \hline
$\zeta_t=\frac{r_t}{r_0}........$ & 29.40\\[3mm]
$\alpha(\zeta_t)=\frac{\bar{p}}{p_t}$... & 3.73\\[3mm]
$\tilde{M}(\zeta_t)..........$ & 61.48\\[3mm]
$b_T=\frac{T_{TIS}}{T_{SUS}}...$ & 2.16\\[3mm]
$\eta=\frac{r_t}{r_m}$........& 0.554\\[3mm]
$\frac{r_{TIS}}{r_{SUS}}............$& 1.11\\[3mm]
$\delta_{L,{\rm cross}}........$&1.52\\[3mm]
$\delta(t_{\rm cross}).......$&97.2\\[3mm]
$\delta(t_{\rm coll}).........$& 130.5\\[3mm]
$\frac{\rho_t}{\rho_0}$................ &$1.95\times 10^{-3}$\\[3mm]
${\cal R}$.................& 514\\[3mm]
$\frac{\rho_0}{\rho_{b,{\rm coll}}}$...........& $1.80\times 10^4$\\[3mm]
$\frac{\rho_t}{\rho_{b,{\rm coll}}}$...........&35.0\\ \hline
\end{tabular}
\end{table}
The mean density inside the TIS is 
\begin{equation}
\label{ave_rho}
\bar{\rho}=3\frac{\tilde{M}(\zeta_t)}{\zeta_t^3}\rho_0
	=7.25\times 10^{-3}\rho_0
=18\pi^2\displaystyle{\left(\frac{b_T}5\right)^3\frac{\tilde{M}_t^3}{\zeta_t^3}
	\rho_{b0}(1+z_{\rm coll})^3}
	=130.5\rho_{b0}(1+z_{\rm coll})^3.
\end{equation}
Together with a numerical solution of the Lane-Emden equation in dimensionless
form for the sphere with parameters given in Table~\ref{summary}, shown in 
Figures~\ref{dens} and \ref{zeta_M_X}, these scaling laws provide 
the complete description
necessary to apply our TIS solution to some cosmological problem.

For some purposes, an approximate fitting formula for the numerical 
solution of the Lane-Emden equation for isothermal spheres, equations
(\ref{nondim_sph}) and (\ref{init_cond}), is convenient. 
Although we have worked  exclusively with the actual numerical 
solution of equation (\ref{nondim_sph}) throughout
this paper, so as to minimize our errors, we shall now provide such a fitting
 formula which is useful as a good approximation 
over a wider range of radii than has been presented in previous treatments in
the literature. We adopt the form
\begin{equation}
\label{rho-analyt_sph}
\tilde{\rho}=\displaystyle{\frac{\rho}{\rho_0}=\frac{A}{a^2+\zeta^2}
    -\frac{B}{b^2+\zeta^2}},
\end{equation}
with the parameters $A$, $B$, $a$, and $b$ sensibly chosen 
(Natarajan and Lynden-Bell 1997; henceforth, NL97).
In the Appendix, we briefly describe the method used to obtain such a 
solution, adjusted to the problem at hand. For the purposes of this paper a
very good approximation is that of equation (\ref{rho-analyt_sph}) with
the following parameters:
\begin{equation}
\label{rho-analyt_our}
(A,a^2,B,b^2)_{TIS}=(21.38,9.08,19.81,14.62),
\end{equation}  
which differs from that in NL97 and is a better fit over a larger range of
radii than theirs (see Appendix A1).
It fits the exact solution within a fractional error of 3 per cent for $0<\zeta<40$,
which is a range that fully encompasses that of our final TIS solution, with
a perfect match to the exact solution at the boundary $\zeta=\zeta_t=29.4$.
\subsection{Comparison of the minimum-energy TIS solution and the uniform
sphere and singular isothermal sphere approximations}
\label{compare_SUS_SIS}

The temperature derived here by the minimum-energy
TIS solution for the postcollapse sphere in virial and hydrostatic
equilibrium which follows top-hat collapse is a factor of approximately two 
larger than the value 
previously derived by satisfying energy conservation and the virial theorem 
for a postcollapse sphere of uniform density for which the surface pressure 
term is neglected, as summarized in \S~\ref{top}. In particular, if we write
$T_{TIS}=b_T T_{SUS}$, then $b_T=\alpha/(\alpha-2)=2.16$. We shall address
the question of the significance of this revision of the postcollapse virial
temperature below. The size of the TIS sphere, given in terms of the radius
$r_m$ of the top-hat at maximum expansion according to $r_t=0.554 r_m$, is
actually not far from the size of the SUS sphere, $r_{\rm vir}=r_m/2$. However, in
terms of the TIS core radius $r_0$, the truncation radius is large, 
$r_t/r_0=29.4$, implying that the TIS is very far from uniform density.   
Despite this relatively small core radius, the TIS 
solution is also quite different from that of a {\it singular}
 isothermal sphere; for the latter, the ratio of the average
density to that at the surface is 3, while we find a value of 3.73 for the
nonsingular TIS, and the
truncation radius in the singular limit is only $r_t=(5/12)r_m$.  
Similarly, the correction factor for the virial temperature of the singular 
isothermal sphere relative to that of the uniform sphere is $b_T=3$, as 
opposed to the value for the TIS which is $b_T=2.16$.  The central density
of the TIS is $1.8\times 10^{4}$ times larger than the mean density
of the background at the collapse epoch for the initial top-hat.  Finally,
this solution predicts that the top-hat will virialize somewhat earlier 
than the nominal collapse time of the top-hat,
since the outermost mass elements encounter a shock in the infall solution at
finite radius.  In terms of the usual overdensity of the top-hat predicted by
linear theory, this implies a revised value of 
$\delta_{c} = (\delta_{c})_{TIS}$
where $(\delta_{c})_{TIS} \approx 1.52$, rather than the standard 
value of 1.686 based upon extrapolating the linear growth to the epoch at 
which the nonlinear top-hat solution predicts infinite density.
\subsection{Comparison of the minimum-energy TIS solution density profile 
with results of numerical simulation of cosmic structure formation
 and with observations}
\label{compare_dens}
The hydrostatic isothermal sphere has been widely adopted as an equilibrium 
model for the density profiles of individual cosmic structures. Mass 
decomposition
studies of galaxies based upon rotation curve data, for example, often adopt 
the 
density profile $\rho\propto [1+(r/a)^2]^{-1}$, an approximation to a 
nonsingular 
isothermal sphere, for the galactic halo (cf. Freeman 1992). As another 
example, galaxy cluster X-ray brightness profiles are often interpreted in 
terms 
of the so-called ``$\beta$-model'' in which the X-ray emitting gas is an 
isothermal sphere in hydrostatic equilibrium in the potential well of a 
virialized
dark matter halo, for the purpose of measuring the cluster total mass density 
and
baryon fraction profiles (e.g. Cavaliere and Fusco-Femiano 1976;
David, Jones and Forman 1995). In this case, the gas density varies as
$\rho\propto [1+(r/a)^2]^{-3\beta/2}$, and the value of $\beta=2/3$, which 
corresponds approximately to that of a nonsingular isothermal sphere, is 
consistent with the data on high temperature X-ray clusters (e.g. Jones \& 
Foreman 1992; Arnaud and Evrard 1998).  These examples of empirical fits are 
only loosely motivated by theoretical expectations for the dynamical evolution
of the objects in question. 

Recently, Navarro, Frenk \& White (1996, 1997; NFW)
 suggested, based upon the results of cosmological N-body simulations of the 
Cold Dark Matter (CDM)
model, that dark matter haloes which condense out of the background can be fit
by a universal density profile, of the following form:
\begin{equation}
\label{NFW}
\displaystyle{\rho_{NFW}=\frac{\rho_S}{(r/r_S)(1+r/r_S)^2}},
\end{equation}
where the fitting parameters $r_S$ and $\rho_S$ are related to each other and 
reflect the different collapse epoch for different mass objects. This NFW
profile shares the large-radius asymptotic shape, $r^{-2}$, of an isothermal
sphere only for intermediate radii ($r/r_S > 1$), while steepening to 
$r^{-3}$ at large radii ($r/r_S>>1$). At small radii ($r/r_S<<1$), the NFW 
profile flattens to $r^{-1}$. Unlike the nonsingular isothermal sphere, 
 the NFW profile lacks a finite core. It is natural to ask, 
therefore, if the apparent differences between the TIS solution presented here
for the postcollapse virial equilibrium of a top-hat density fluctuation and 
the NFW density profile are significant and, if so, what they mean. A full 
treatment of this question is outside of the scope of this paper. However, we 
will briefly address the issue as follows.

There are several questions which must be answered in order to compare the 
TIS and NFW profiles in a meaningful way. First, are they solutions of the same
initial value problem? Second, is the NFW halo a state of relaxed, virial 
equilibrium as the TIS solution is posited to be? Third, is the NFW profile 
really an accurate fit to the exact results for the dark matter haloes which 
result from the growth of density fluctuations in a CDM model universe. 
Fourth, is there a difference between the gas and dark matter profiles 
expected when objects condense out in the CDM model in general, or when the 
dark matter profile follows a NFW shape, in particular? 
Fifth, what do the observations of actual density
profiles for cosmologically condensed, virialized structures say regarding the
comparison between the TIS solution presented here and the NFW profile?

Regarding the first question, the NFW profile is suggested to be the generic
outcome of the condensation of haloes of collisionless dark matter in 
hierarchical
clustering models like CDM, starting from Gaussian random noise density 
fluctuations. The TIS profile presented here is the outcome of the collapse of
 an
uncompensated spherical top-hat, instead, albeit one in which the presence of
some smaller-scale nonuniformity leads to relaxation to a final equilibrium 
state
of finite density. We have argued that a solution very similar to our minimum
energy TIS also arises if we consider the self-similar spherical infall 
solution
of Bertschinger (1985), which is already close to isothermal and hydrostatic 
equilibrium inside the shocked region, and assume that rapid equilibration 
takes
place within this shocked region. In that case, there is a continuous infall 
which leads to a continuous sequence of TIS solutions of ever larger mass as 
more
mass encounters the accretion shock over time, and the shock represents a 
boundary
which separates the relaxed region in equilibrium inside the shock from the
unrelaxed, infalling matter at larger radii. As such, the TIS and NFW profiles
do not necessarily arise from identical initial conditions.

The simple spherical top-hat model has been quite successful, however, in
characterizing the rate of formation of haloes of different mass at different 
times in hierarchical clustering models, as in the Press-Schechter 
approximation.
The standard approach to this problem is to filter the density field in such 
models spatially with a spherical window function and to apply the top-hat
model to predict the nonlinear collapse of initially linear-amplitude density
fluctuations whose wavelength corresponds to the size of the window function.
 In a similar way, we might expect the TIS solution found here to be useful in 
characterizing the internal structure of the virialized objects which result 
from this nonlinear collapse. We note that there is currently no fundamental
derivation or detailed theoretical explanation of the empirical NFW profile,
so it is difficult to say just which initial conditions are necessary to 
produce it and how different these can be from that of the top-hat or of the 
spherical infall solution.

Some attempt to address this was made recently by Huss, Jain and Steinmetz 
(1998), who added different amounts of velocity dispersion to the initial 
conditions of a spherically symmetric density perturbation evolved by N-body
simulation to see if the NFW profile would still fit the end result of its 
collapse. They found that the NFW was a decent fit for a range of initial 
velocity dispersions, but that it was possible to fit the simulation results
better with a profile that was flatter at small radii than the NFW profile, as
flat as $r^{-0.5}$. Syer and White (1998), on the other hand, have argued that 
the actual halo profiles in hierarchical clustering models are the product of
the merging history of subclumps to form the halo, as an explanation for the 
cuspy inner density profile. They suggest that the slope of this inner density 
profile is different for haloes arising from linear density fluctuations whose
initial power spectra have different slopes. If so, then the NFW is not a 
unique fit, since the actual profiles must depend upon the slope of the power
spectrum at the wavelengths responsible for producing haloes of a given mass.
This suggests that the haloes observed in N-body simulations are not fully 
relaxed,
in the sense that they have not lost memory of the initial inhomogeneities 
which led to their formation. According to Syer and White (1998), in fact, the
faster the initial density fluctuation power spectrum $P(k)\propto k^n$ 
decreases with increasing $k$, the shallower is the predicted central density 
profile of the haloes which form on this scale, with 
$\rho\propto r^{-\gamma}$,
and $\gamma\approx 3(3+n)/(5+n)$. For the particular case of galactic scales
in the standard CDM model, the effective power-law index is $n\approx -2.5$ and
the scaling law would predict $\gamma\approx 0.6$, smaller than the NFW value.
Only for larger-mass haloes, those at galaxy cluster scales, where 
$n\approx -2$,
does this argument predict that $\gamma\approx 1$ as in the NFW profile. We 
note
that, according to the scaling law of Syer and White (1998), as $n$ approaches
$-3$, the approximate asymptotic value in the small wavelength limit in the CDM
model, the inner profile is predicted to flatten to a constant core value. It 
is tempting to speculate that the scaling argument, if correct, might be 
pointing to
our TIS solution for the virialized inner portion of a halo which condenses out
of a CDM model universe, for smaller mass haloes, those 
which originate from density 
fluctuations in the small wavelength limit. In this limit, there are mass 
fluctuations of comparable amplitude on all scales, collapsing out nearly 
simultaneously, which may be the limit in which the phenomenon of repeated
mergers of isolated, dense subclumps to build up a larger halo is not prevalent
enough to generate a central density cusp.

We have focused above on the differences between the NFW and TIS profile 
behaviour
at small radii, but there is also a difference at large radii; the logarithmic 
slope of the TIS density profile drops as low as $-2.5$ at intermediate radii
($r/r_0\approx 9$) and rises to $-2.1$ at the outer radius (see 
Fig.\ref{dens}), whereas the outer slope of the NFW profile approaches $-3$ 
(i.e. at the outer radius, $(r/r_S)_{max}$, the NFW slope is 
$3-2/((r/r_S)_{max}+1)$).
 This difference may reflect the transition between the inner virialized
part of the halo, which we have characterized as isothermal and hydrostatic
in formulating our TIS solution, and the outer region, involving continuous
infall, which is neither isothermal, nor hydrostatic. The infall, itself, is
generally predicted by spherical models to have a density profile closer to 
$r^{-2}$ than to $r^{-3}$ (e.g. Gunn and Gott 1972; Fillmore and Goldreich 
1984;
Hoffman and Shaham 1985; Bertschinger 1985; Moutarde, et al. 1995). When
account is taken of the presence of an accretion shock in the fluid case or of
density caustics and shell-crossing in the case of collisionless matter, such 
as the self-similar infall solution of Bertschinger (1985), the density profile
{\it inside} the shocked region and the region of shell-crossing is also closer
to $r^{-2}$ than to $r^{-3}$, even before any dynamical relaxation of the 
matter
to an equilibrium profile occurs. Across the boundary between these two 
regions,
the outer region of infalling matter and the inner region of shocked fluid and
collisionless shell-crossing, however, there is a density jump, with a higher 
density inside the outermost density caustic than just outside this radius. 
To the extent that such a spherical model can represent the average behaviour 
of a more inhomogeneous initial condition, therefore, we expect the average 
density in the more general case to drop more steeply than $r^{-2}$ across 
some transition zone at $r\sim r_{\rm vir}$ where the actual transition in the 
collisionless matter is not
discontinuous but is rather of finite radial extend. This would account 
qualitatively for the steepening in the NFW profile from $r^{-2}$ to $r^{-3}$
at large radii.

The question of whether the haloes in the N-body simulations of the CDM model,
fit by the NFW profile, are relaxed and in equilibrium has been partially 
addressed by Tormen, Bouchet and White (1998) for the case of collisionless 
matter in an Einstein-de Sitter universe in which the density fluctuations are
Gaussian random noise with a power spectrum which is a scale-free power-law
of index $n=-1$. They report that the haloes, while not isothermal, are not very
far from isothermal. They also find that the particle velocities at small
radii are nearly isotropic while those at large radii are predominantly radial.
As another approach, these authors compared different mass estimators for a
set of haloes, all based upon an application of the Jeans equation for the
dynamical equilibrium of a spherically symmetric collisionless system, allowing
for different degrees of departure from the condition of velocity dispersion
isothermality and isotropicity. They found that these mass estimators all
compared favorably with the true mass if applied to estimate the total mass
interior to the virial radius, while the estimator which took account of 
anisotropic velocity dispersion was in good agreement with the true mass at all
radii. This suggests that the haloes are approximately and on average in a state
of dynamical equilibrium but not fully relaxed to the point of having isotropic
and isothermal velocity dispersion. As an additional probe, Tormen et al.
(1997) also considered the fate of their simulated haloes at late times under
the artificial circumstance in which a halo identified at some epoch is 
allowed to evolve further but with all matter removed outside of a sphere of
radius twice the size of the original halo. This was intended to break the
clustering hierarchy so as to allow the halo to evolve to a relaxed 
equilibrium state in the absence of further influx of perturbing substructure.
Interestingly, the density profiles of the resulting haloes tended to flatten
thereafter at very small radii, although the authors suggested that this might
reflect some numerical error or artifact. Otherwise, the haloes showed less
substructure with time but continued to be well fit, outside the central 
region, by the NFW profile.
 
What do numerical N-body simulations tell us about the accuracy or universality
of the NFW profile for the haloes that form in the particular Gaussian random
noise, hierarchical clustering model, CDM, which it was originally proposed to
fit? Unfortunately, the results to date are not yet conclusive.
For the purposes of discussion, it is useful to  consider a broader family of
density profiles, of the following form:
\begin{equation}
\label{gen_profile}
\displaystyle{\rho(r)=\frac{\rho_0}{\left(\frac{r}{r_S}\right)^\gamma
 \left(1+\left(\frac{r}{r_S}\right)^{1/\alpha}\right)^{\alpha(\beta-\gamma)}}}
\end{equation}
(e.g. Hernquist 1990; Zhao 1996). This density profile approaches a power law of slope 
$-\gamma$ at small radii, and a power law of slope $-\beta$ at large radii, 
with a break at the characteristic radius $r_S$ and a transition region between
whose width increases with increasing $\alpha$. The NFW profile corresponds to
the parameters $(\alpha,\beta,\gamma)=(1,3,1)$. An earlier suggestion of a 
galactic halo density profile for elliptical galaxies by Hernquist (1990)
corresponds to $(\alpha,\beta,\gamma)=(1,4,1)$. A comprehensive review of the
attempts to fit numerical simulation results for halo density profiles to 
those of the form in equation (\ref{gen_profile}) above is beyond the scope of 
the present paper. We will mention only a few of the most recent and refer the
reader to these papers and references therein for further description. Tormen
{et al.} (1997) compare their simulation results by an N-body tree-code for
haloes in the case of a scale-free power-law power spectrum of initial density
fluctuations with $n=-1$ to the NFW and Hernquist profiles, resolving 
individual haloes with $20,000$ simulation particles. Fitting the region 
between $0.01$ and one times the virial radius (defined roughly in terms of the
conventional SUS approximation for the average overdensity within a 
postcollapse, virialized top-hat), they find a somewhat better fit for the NFW
profile than for the Hernquist profile. Recent high-resolution simulations of 
galaxy halo formation in the CDM model by Kravtsov, {et al.} (1998), 
however, by a new Adaptive Refinement Tree (ART) N-body code with a comparable
number of particles per halo and a larger sample of haloes to analyze find a 
systematic deviation from the NFW profile at small radii, with a much flatter
central profile, corresponding to $\rho(r)\propto r^{-\gamma}$ with $\gamma 
\approx 0.2$ (i.e. $(\alpha,\beta,\gamma)\approx(0.5,3,0.2)$; we note that
Kravtsov, {et al.} define their own $\alpha$ as the inverse of the one
we use here, so we transform their value to our notation). At small radii,
at least, this is perhaps more consistent with the flat core of the TIS 
solution than with the cuspy center of the NFW 
density profile. On the other hand, very high resolution N-body simulations
by a parallel N-body tree-code, reported by Moore, {et al.} (1998), of two
rich-cluster-sized haloes in a CDM model in which a range of resolutions up to 
more than $10^6$ particles within the virial radius was used, finds that the
simulated profiles deviate at small radii from NFW in the opposite sense, 
becoming as steep as $\rho(r)\propto r^{-\gamma}$ with $\gamma \approx 1.4$
(i.e. $(\alpha,\beta,\gamma)\approx(0.7,2.8,1.4)$), which they believe is
reliably determined down to radii less than $0.01$ times the virial radius.
A qualitatively similar result regarding the density cusp is reported by 
Fukushige and Makino (1998) based upon their N-body simulations of the 
formation of a single, massive, galactic halo in the CDM model by direct
summation of two-body gravitational forces utilizing the Grape-4 
special-purpose computer. The full simulation volume of comoving radius 2 Mpc
in present units contained 786,000 particles and more than $10^{12} M_\odot$,
but they do not indicate what fraction of these particles were contained within
the virial radius of the final object.

There are at least two possible explanations for the different results of these
latest simulations. It may be that, as Syer and White (1998) suggest, the NFW
profile is not universal but depends instead on the shape of the density 
fluctuation power spectrum at wavelengths responsible for the particular haloes
in question. As such, the galactic halo profiles of Kravtsov, {et al.} are
expected to be flatter at small radii than NFW, as described above, in contrast
to the large, cluster-mass haloes of Moore, {et al.} (1998). This scaling
argument alone does not, however, explain why the Moore {et al.} (1998)
haloes are more cuspy even than those of Tormen et al. (1997) for the
scale-free $n=-1$ case, since, although the effective $n$ in the CDM model 
does increase from galactic to cluster mass scales, it does not exceed $-1$ on
the cluster scale, so the  Tormen {et al.} (1997) results, too, should show
a greater degree of central cuspiness than does NFW and they do not. Nor does 
this explain why the simulations of Fukushige and Makino (1998) of a
galactic halo, rather than of a cluster halo, also show a density cusp steeper
than the NFW profile. A second
possibility is that, despite the improvement in mass resolution afforded by
these new simulations, numerical effects are still responsible for a failure
to resolve the inner regions of the haloes properly. Splinter {et al.} 
(1998) for example, have raised a significant question about whether 
artificial collisional relaxation effects in N-body codes used to model 
collisionless matter have been underestimated, suggesting that previous claims
of length resolution below the mean interparticle separation are not correct.
Further work will be necessary to resolve this matter.

Regarding the question of the possible difference between the gas and 
collisionless dark matter distributions, we have assumed this difference to be 
relatively small in our TIS derivation, so that our TIS solution can be applied
to either component, in proportion to its relative share of the total mean 
matter density in the unperturbed background universe. This was motivated, in 
part, by the self-similar spherical infall solution of 
Bertschinger\shortcite{Bert85} which supports this assumption if the gas has a
ratio of specific heats $\gamma=5/3$. It is also supported by coupled N-body
and gas dynamical simulations of the CDM model (i.e. those which neglect both
radiative cooling and external input of energy to the gas, such as by
photoionization and supernova explosions). For example, cluster simulations by
Eke, Navarro, and Frenk (1998) find that the cumulative gas mass fraction 
measured from the cluster center (in units of the universal mean value 
over all space) is unity outside the virial radius, roughly $0.9$ between
$0.5 r_{\rm vir}$ and $r_{\rm vir}$ and drops below this significantly only in the 
core. Interestingly, while the density profile of the cluster dark matter 
in those simulations is well-fit by the NFW profile, the gas density profile 
does depart somewhat in the center, where it is found to possess a finite core,
and it is better-fit overall by a $\beta-$model like that traditionally 
used to model cluster 
X-ray gas as a hydrostatic isothermal sphere, with $\beta$ between $0.7$ and
$0.75$ for the average profiles at different epochs, closer to our TIS 
solution than to the NFW profile of the dark matter. The ratio of the virial 
radius to the core radius of the average cluster density profile at $z=0$, in
fact, was found to be $20$, as compared to our TIS solution in which the 
ratio of truncation radius to core radius is $29.4$. (Note: our truncation 
radius is, moreover, a little higher than the SUS approximation for $r_{\rm vir}$,
so if we replace $r_t$ by $r_{\rm vir}=0.5r_m$, our ratio drops to $26$, even 
closer). Bryan and Norman\shortcite{B&N} report that their simulations of
cluster formation in the CDM model by a different numerical technique also
support the assumption that the cluster gas density profile follows the dark
matter density profile throughout most of the cluster volume, with differences
important only at $r/r_{\rm vir}< 0.04$. In short, our approximation that the same 
density profile can be used to describe both gas and dark matter in the 
postcollapse virial equilibrium of a top-hat density perturbation seems 
justified, in view of these simulation results and the uncertainty described 
above in 
determining the true nature of the dark matter profiles at small radii (i.e.
radii as small or smaller than the core radius of our TIS solution). 

We note that, while the dark matter in these 
simulated haloes in the CDM model is generally
found to exhibit velocity dispersion which is somewhat radially-biased outside
the core, the bias is relatively modest except at radii close to $r_{\rm vir}$. In 
particular, if $\beta_{an}\equiv 1-\bar{v}_t^2/2\bar{v}_r^2$, where $v_t(v_r)$
correspond to tangential (radial) velocities, then Eke 
{et al.} (1998) report $\beta_{an}<0.2$ out to $r=r_{\rm vir}/2$, rising only
to 0.4 at $r\approx r_{\rm vir}$. The gas, of course, cannot have an anisotropic 
temperature but could share the radial bias in the sense of there being
a net infall or outflow, if it were not hydrostatic. The radial velocity
bias of the 
dark matter in the outskirts of the cluster need not be inconsistent with our
assumption that the gas follows the dark matter as long as it reflects the
transition from infall at larger radii to hydrostatic equilibrium at smaller
radii for both the gas and the dark matter, as already described above to
explain the steeper $r^{-3}$ dependence of the NFW profile at these radii, 
compared to $r^{-2}$ for the TIS solution. The ratio of dark matter velocity
dispersion to gas temperature is reported to be close to unity throughout the
simulated clusters, except at the very center, where the gas remains isothermal
but the dark matter ``temperature'' drops, inside the gas density core radius,
in the inner region where the cuspy density of dark matter is claimed to occur.
In short, there are many indications from these simulations that our TIS 
solution is a plausible approximation for the behaviour of the gas, and also
some indication from the similarity between the gas and dark matter outside 
the gas density core that our assumption that gas and dark matter follow the 
same solution is valid.

Finally, we return to the question of whether the NFW profile for the dark 
matter in haloes that form in the CDM model is in agreement with the 
{\it observed} properties of dark matter haloes in nature. A full treatment of 
this question is outside the scope of our discussion. We focus, instead on a 
few recent attempts to answer this question on two widely different scales.
The first involves dwarf and low-surface-brightness (LSB) late-type galaxies.
Such systems are believed to be dark-matter dominated, and observations of 
their rotation curves have been used to probe the density structure of dark 
matter haloes directly. Flores and Primack\shortcite{F&P} and 
Moore\shortcite{Moore} used rotation curve measurements for several such 
galaxies and concluded that the central density distribution could not be
as cuspy as the $r^{-1}$ shape of a NFW profile. Burkert\shortcite{Bur} showed
that the data of Moore\shortcite{Moore} was well fit by a profile of the form
\begin{equation}
\label{bur_profile}
\displaystyle{\rho(r)
	=\frac{\rho_0}{\left(1+\frac r{r_S}\right)
	\left[1+\left(\frac r{r_S}\right)^2\right]}},
\end{equation}
which differs significantly from the NFW profile in that it exhibits a finite 
core. Kravtsov {et al.}\shortcite{Kravtsov} have examined this question 
with a somewhat larger observed sample of dwarf and LSB galaxy rotation curves
and concluded that a good fit to these halo density profiles of the form in 
equation (\ref{gen_profile}) involves a shallow central profile 
$\rho\propto r^{-\gamma}$, with $\gamma=0.2-0.4$, shallower than the NFW 
profile. Burkert and Silk\shortcite{B&S} reexamined the rotation curve of one
dwarf galaxy, DDO154, to suggest that the discrepancy with the NFW profile 
could be resolved if a spheroid of dark {\it baryons} were added to the CDM 
halo of comparable mass. In short, on galaxy scales, the well-studied systems 
which are considered to be the most direct measure of the dark matter halo are
found to be inconsistent with the NFW in requiring either a finite core or a
shallow inner profile. This is the same sense as the discrepancy between the
TIS and NFW profiles.

On the larger scale of clusters of galaxies, the observations of the X-ray
emission by the gas and of gravitational lensing by the dark matter provide two
independent measures of the distribution of the dark matter in the cluster.
Makino, Sasaki and Suto\shortcite{MSS} found that the gas density profile of 
an X-ray cluster, if modeled as an isothermal sphere in hydrostatic 
equilibrium with a dark matter potential well which follows the NFW density
distribution, yields an isothermal $\beta-$profile with a finite core, as
required to match the observations of X-ray cluster brightness profiles, but
with too small a core radius relative to the observed ones. Eke 
{ et al.}\shortcite{ENF} performed SPH simulations of the formation of such
clusters in a CDM model and claim that their cluster dark matter follows the
NFW profile while the gas density profiles are well fit by the isothermal 
$\beta-$model, with a typical core radius for the latter of about $100\,h^{-1}$
kpc at $z=0$, in a good agreement with observations. The origin of the 
difference between these two conclusions is not clear.

An independent comparison of the dark matter density profiles in observed 
clusters of galaxies and the NFW profile is provided by a recent observation 
of strong gravitational lensing of a background galaxy by the galaxy cluster
CL 0024+1654 at $z=0.39$ by Tyson, Kochanski, and Dell'Antonio\shortcite{TKD}.
These authors construct a high-resolution map of the central mass distribution
of the cluster and find it to be inconsistent with the singular density profile
of NFW. The observed cluster requires a smooth dark matter distribution with
a $35\,h^{-1}$ kpc core and a slope that is slightly shallower than that of an
isothermal sphere. This is consistent with the TIS solution presented here.

\subsection{Comparison of the predicted virial temperature and
mass -- radius -- temperature scaling laws of the minimum-energy TIS solution 
with numerical simulation results and observations of X-ray clusters}
\label{compare_temp}

We have noted above that our minimum-energy TIS solution temperature is 
larger by a factor of 
$b_T=2.16$ than the postcollapse virial temperature calculated in the SUS
approximation, while it is smaller by a factor of $3/2.16= 1.39$ than
the virial temperature in the limit of a {\it singular} isothermal sphere.
How does the TIS temperature compare with the results of numerical 
simulations of cosmological structure formation? The clearest comparison 
available to date is with the results of numerical gas dynamics simulations of 
X-ray cluster formation in the CDM model. It is customary to describe these 
results by comparing the numerical gas temperatures with the so-called 
``virial temperature'' $T_{\rm vir}$ of the clusters. Unfortunately, different
authors define their $T_{\rm vir}$ differently, often without explicit 
justification, which makes a direct comparison more difficult. We shall 
attempt to compare the results quoted in the literature uniformly by 
referring to the definition of virial temperature used by Kitayama and
Suto\shortcite{Kitayama_Suto}, given by
\begin{equation}
\label{fudge}
k_BT_{\rm vir}\equiv \gamma_{\rm vir}\frac{G\mu m_p M}{3r_{\rm vir}},
\end{equation}
in which $r_{\rm vir}$ is the virial radius, which different authors have chosen 
to define in different ways when analyzing their simulations (e.g. sometimes
$r_{\rm vir}$ is set equal to $r_{200}$, the radius within which the average 
density is $200$ times the mean background value), and $\gamma_{\rm vir}$ is
a ``fudge factor'' which typically ranges between $\gamma_{\rm vir}\approx 1$ and 
$\gamma_{\rm vir}\approx 1.5$ and is chosen either to match simulation results or 
according to some analytical model like the SUS approximation.
Kitayama and Suto\shortcite{Kitayama_Suto}, for example, adopted the value
 $\gamma_{\rm vir}=1.2$ based upon the gas dynamical simulations of White {et al.}
(1993),  while pointing out that other authors adopted other values in their
analyses, including $\gamma_{\rm vir}=1$ \cite{Kitayama_Suto0},  $\gamma_{\rm vir}=1.1$ 
\cite{V&L}, and $\gamma_{\rm vir}=1.5$ (Eke {et al.} 1996). A complete review of
which authors adopt which value of $\gamma_{\rm vir}$ in defining $T_{\rm vir}$ is outside 
the scope of this paper. We shall focus, instead, on some recent papers for 
which simulation results can be used to determine the value of $\gamma_{\rm vir}$.

We note that the same expression used in equation (\ref{fudge}) can be used to
define the virial velocity dispersion, $\sigma^2$, of the collisionless dark 
matter component, according to equation (\ref{dispersion}), if we replace
$k_BT_{\rm vir}/(\mu m_p)$ in equation (\ref{fudge}) by $\sigma^2$. Henceforth
we shall distinguish the $\gamma_{\rm vir}-$values for the gas and the dark matter 
by the subscripts ``$T$'' for the gas and ``$\sigma$'' for the dark matter, and
drop the subscript ``vir'' for simplicity. Our 
assumption in this paper is that the gas and dark matter components each 
follow the 
same TIS profile with component densities which are simply proportional to
the total density at each point, and that the two components have the same 
virial temperature or equivalent virial velocity dispersion $\sigma^2$. The
latter property is sometimes quantified by defining a parameter 
$\beta_{\sigma-T}\equiv \mu m_p \sigma^2/k_B T$ (i.e. 
$\beta_{\sigma-T}=\gamma_{\sigma}/\gamma_T$).
We have assumed here that $\beta_{\sigma-T}=1$ is a good approximation. 

A comparison of our TIS solution result, the SUS and SIS approximations, and 
the results of some of the more recent numerical simulation studies of cluster
formation in the CDM model is presented in Table~\ref{M-T}. For the TIS 
solution evaluated at the boundary $\zeta=\zeta_t$, 
$\gamma_T=\gamma_{\sigma}=(6/5)(b_T\eta)=1.43$ and $\beta_{\sigma-T} = 1$, 
while the SUS approximation corresponds to $\gamma_T=\gamma_{\sigma}=0.6$ and 
$\beta_{\sigma-T}=1$, and the SIS approximation yields 
$\gamma_T=\gamma_{\sigma}=1.5$ and $\beta_{\sigma-T}=1$.
\begin{table*}
\centering
\begin{minipage}{140mm}
\caption{Comparison of the minimum-energy TIS solution and 
X-ray cluster simulation results for the mass-temperature virial relation}
\label{M-T}
\begin{tabular}{@{}llcrc}
&Source&$\gamma_T$& $\gamma_{\sigma}$&$\beta_{\sigma-T}$\\ \hline
&Simulation Results&&\\  \hline
&Eke, Navarro, and Frenk (1998)...........&1.5&1.5&1\\ 
&Bryan and Norman (1998)...................& 1.2&1.2&1\\ 
&Evrard, Metzler, and Navarro (1996)...&1.4&1.4&1\\
&Crone and Geller (1995).......................&--&1.2&--\\
&Cole and Lacy (1996)...........................&--&$\approx 1.5$&--\\ 
		\hline
&Analytical Results&&\\  \hline
&Uniform Sphere (``SUS'')............................................&0.6&0.6&1\\
&Singular Isothermal Sphere (``SIS'')............................&1.5&1.5&1\\
&Truncated Isothermal Sphere (``TIS'', this paper)
\footnote{For the TIS solution, $\gamma_{\rm vir}$ is evaluated at
the outer boundary, $\zeta=\zeta_t$. The values for other radii are 
plotted in Figure~\ref{zeta_M_X}, showing that $\gamma_{\rm vir}$ varies
roughly between $1.2$ and $1.5$ for radii between $0$ and 
$r_t$.}... &1.43&1.43&1\\ \hline
\end{tabular}
\end{minipage}
\end{table*}
The numerical gas dynamical simulations all report $\beta_{\sigma-T}\approx 1$,
with
$\gamma_T$ and $\gamma_{\sigma}$ ranging between $1.2$ and $1.5$. The purely
collisionless $N$-body simulations (without gas dynamics) also find 
$\gamma_{\sigma}$ between $1.2$ and $1.5$. These results are in good agreement
with the prediction of our TIS solution, in rough agreement with the SIS
approximation and in clear {\it disagreement} with the SUS
approximation. At a more detailed level, there are some disagreements between
the TIS solution and the simulation results and amongst the simulation results
themselves. For example, Eke {et al.} (1998) report that the region of
hydrostatic and virial equilibrium is most reliably located at 
$r\simeq r_{\rm vir}/3$ and Evrard{ et.al.} (1996) (hereafter EMN96)
note a similar trend, 
although departures from equilibrium are never very great even at $r_{\rm vir}$.
Similarly, while the clusters are found in all cases to be approximately
isothermal, there is some tendency for the temperature to fall by less than a 
factor of two roughly between $r_{\rm vir}$/2 and $r_{\rm vir}$. Bryan and 
Norman\shortcite{B&N} find a small tendency for $\beta_{\sigma-T}$ to decrease
with 
cluster mass, smaller than unity for smaller clusters, but approaching unity 
for the larger clusters, which they attribute to the effect of numerical
resolution in better resolving the more massive clusters. Finally, Eke 
{et al.} (1998) find a small drop in the dark matter $\sigma^2$
relative to the gas temperature $T$ near the center, consistent with
their finding that the central density of the dark matter is more centrally 
peaked than is that of the gas. Some of the differences amongst the simulation
results may reflect a weak dependence of those results on the different 
background cosmology models adopted. Nevertheless, within the uncertainties 
of the 
simulation results, there is good global agreement between the TIS solution 
presented here and the numerical simulations of cluster formation in the CDM 
model in terms of the temperature -- mass -- radius virial relation. 

EMN96 have used a large set of simulated clusters to 
infer a mass-temperature scaling law and a related radius-temperature scaling 
law for the clusters. While the functional dependences of mass and radius on 
temperature in these scaling laws are consistent with the crude expectations
of virial equilibrium as measured at a fixed value of the average density 
contrast inferior to some radius $r$ within the cluster relative to the 
mean background density $\rho_b$, $\bar{\rho}/\rho_b$, namely,
 $T\sim GM/r\sim (\bar{\rho}/\rho_b)\rho_br^2$, the coefficients of the
scaling laws are determined empirically from the numerical gas dynamical 
simulations of X-ray cluster formation in the CDM model. By making a 
conservative choice, $\rho/\rho_b=500$, of an average density contrast which 
is generally expected to occur at a radius which is significantly smaller than 
that within which the cluster
is well-approximated by hydrostatic and virial equilibrium, EMN96
 are able to calibrate these scaling laws 
numerically, with relatively small scatter, with the intention of comparing 
them with X-ray observations of clusters to determine reliable mass estimates
for the clusters. For our purposes, these numerically -- calibrated scaling
laws provide an excellent check on our analytical TIS model.
If our TIS model solution proves to be in good agreement with these scaling
laws, then we will not only have confirmed our TIS analytical predictions
but also have found in our TIS solution a {\it derivation} of these
numerically -- calibrated scaling laws.

Let $r_{500}$ be the radius of the sphere within which the mean density 
contrast is $\bar{\rho}/\rho_b=500$, and let $M_{500}$ be the mass enclosed 
by this sphere.  EMN96 find 
\begin{eqnarray}
\label{M500}
M_{500}(T)
&=&\displaystyle{(1.11\pm 0.16)\times10^{15}\left(\frac T{10\, {\rm keV}}
	\right)^{3/2}h^{-1}\, M_\odot},\\
\label{r500}
r_{500}(T)
&=&\displaystyle{(1.24\pm 0.09)\left(\frac T{10\, {\rm keV}}\right)^{1/2}
		h^{-1}\, {\rm Mpc}},
\end{eqnarray}
where $T$ is an appropriate average temperature for the cluster gas. The 
simulation results which led to equations (\ref{M500}) and (\ref{r500})
were analyzed at $z=0.04$, but scaled to $z=0$ assuming that they both
scale as $(1+z)^{-3/2}$, so there is no information explicitly contained in 
these equations regarding the scaling with redshift. The corresponding scaling
laws for a mean density contrast $\bar{\rho}/\rho_b=200$, were presented in 
Arnaud and Evrard \shortcite{A&E} (with no error bars specified, but the error
bars should be larger than those at smaller radii; Evrard, private 
communication) for the same 
numerical simulation results as in EMN96, according to:
\begin{eqnarray}
\label{M200}
M_{200}(T)
&=&\displaystyle{1.45\times10^{15}\left(\frac T{10\, {\rm keV}}
	\right)^{3/2}h^{-1}\, M_\odot},\\
\label{r200}
r_{200}(T)
&=&\displaystyle{1.85\left(\frac T{10\, {\rm keV}}\right)^{1/2}
		h^{-1}\, {\rm Mpc}}.
\end{eqnarray}

We can derive the corresponding relations according to our minimum-energy TIS 
solution. Let $X$ be the average density contrast inside
a sphere of radius $r_X$ which encompasses a total mass $M_X$.
Then the minimum-energy TIS solution derived here can 
be used to solve for $M_X$ and $r_X$ for the postcollapse virialized
object of virial temperature $T$ which results from a top-hat which collapses
at any $z_{\rm coll}$, for any overdensity $X$, as follows. 
Let $\bar{\rho}_X$
be the average density inside this sphere of radius $r_X$. In terms of 
$\rho_b(t_{\rm coll})$, the mean cosmic background density at the time $t_{\rm coll}$ 
for the top-hat perturbation which created the postcollapse virialized object
described by the minimum-energy TIS solution, the density contrast is
$X=\bar{\rho}_X/\rho_b(t_{\rm coll})$. We shall continue to refer to the values of 
these quantities at the outer radius $r_t$ by using the subscript ``t'' as we have
in previous sections. In terms of the dimensionless radius $\zeta$ and 
mass $\tilde{M}$, the fact that $\bar{\rho}_X\propto M_X/r_X^3$ allows us to
write
\begin{equation}
\label{ratio}
\frac{\tilde{M}_X/\zeta_X^3}{\tilde{M}_t/\zeta_t^3}
	=\frac X{\bar{\rho}_t/\rho_b(t_{\rm coll})}.
\end{equation}

\begin{figure}
\centerline{\psfig{figure=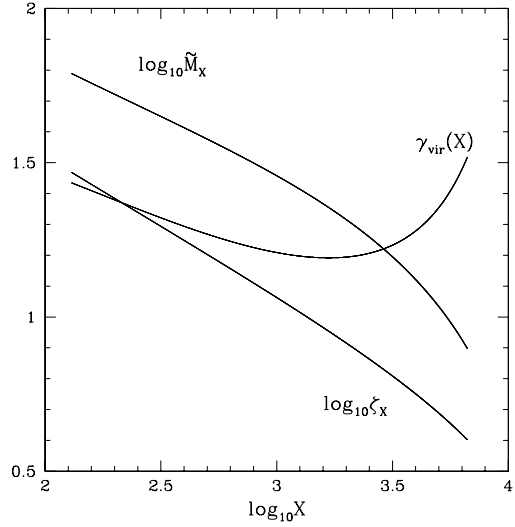,height=3in,width=3in}}
\caption{The dimensionless radius $\zeta_X$, and the corresponding mass 
$\tilde{M}_X$, within which the average density with respect to the background
is $X$, and the factor $\gamma_{\rm vir}$, as functions of the density $X$ for
our minimum-energy TIS solution.}
\label{zeta_M_X}
\end{figure}

Equations (\ref{ave_rho}) and (\ref{ratio}) combine to yield an implicit equation
for $\zeta_X$ and, hence, for $\tilde{M}_X$, which can be solved numerically for
any given $X$, according to
\begin{equation}
\label{implicit}
\displaystyle{\frac{X\tilde{\rho}_X^3}{\tilde{M}_X}
	=18\pi^2\left(\frac{b_T}5\right)^3\tilde{M}_t^2}.
\end{equation}
The mass $M_X$ can be expressed in terms of the resulting $\tilde{M}_X$, using 
equations (\ref{mass_nond}), (\ref{r_0_final}), and (\ref{rho_0_final}),
 according to
\begin{equation}
\label{M_X}
\displaystyle{\frac{M_X}{\tilde{M}_X}
	=\frac 1{3\pi G}\left(\frac{b_T}5\right)^{-3/2}\tilde{M}_t^{-1}
	(1+z_{\rm coll})^{-3/2}H_0^{-1}\left(\frac{k_BT}{\mu m_p}\right)^{3/2}}.
\end{equation}
Similarly, the radius $r_X$ can be expressed in terms of the dimensionless 
radius $\zeta_X$ which is the solution of equation (\ref{implicit}) and the
core radius $r_0$, according to
\begin{equation}
\label{r_X}
\displaystyle{\frac{r_X}{\zeta_X}=r_0
	=G\frac{\mu m_p}{k_B T}\frac{M_X}{\tilde{M}_X}
	=\frac 1{3\pi}\left(\frac{b_T}5\right)^{-3/2}\tilde{M}_t^{-1}
	(1+z_{\rm coll})^{-3/2}H_0^{-1}\left(\frac{k_BT}{\mu m_p}\right)^{1/2}},
\end{equation}
where in expressing $r_0$ we used equations (\ref{r_0}) and (\ref{mass_nond}).
Equations (\ref{implicit}), (\ref{M_X}), and  (\ref{r_X}) constitute the final 
scaling laws for the mass and radius within our TIS for the interior spherical 
region of average density contrast $X$, in terms of the predicted virial 
temperature $T$, collapse epoch $z_{\rm coll}$ and Hubble constant $H_0$, if we 
use the results of our TIS solution to evaluate $b_T$ and $\tilde{M}_t$. 
Plots of $\zeta_X$ and $\tilde{M}_X$ as functions of $X$, obtained numerically
from the TIS solution, are shown on Fig.~\ref{zeta_M_X}, along with a plot of 
$\gamma_{\rm vir}(X)\equiv 3\tilde{M}_X/\zeta_X$ which enters equation 
(\ref{fudge}) if we wish to evaluate equation (\ref{fudge}) using any radius
$r\leq r_t$ and replace $M$ by $M(\leq r)$ and $r_{\rm vir}$ by $r$.

In order to compare directly with the empirical scaling laws of EMN96
in equations (\ref{M500}), (\ref{r500}), (\ref{M200}), and (\ref{r200}) above,
we set $X=500$ or $X=200$, accordingly, and use our
minimum-energy TIS solution summarized in Table~\ref{summary} to evaluate 
equations (\ref{implicit}) -- (\ref{r_X}). We find $\zeta_{500}=15.96$,
$\tilde{M}_{500}=37.74$, $\zeta_{200}=24.20$, and $\tilde{M}_{200}=52.65$. 
The analytical TIS scaling laws which result are the following
\begin{eqnarray}
\label{M500_our}
M_{500}(T)
&=&\displaystyle{1.11\times10^{15}\left(\frac T{10\, {\rm keV}}\right)^{3/2}
    (1+z_{\rm coll})^{-3/2}\left(\frac{\mu}{0.59}\right)^{-3/2}h^{-1}\, M_\odot},\\
\label{r500_our}
r_{500}(T)
&=&\displaystyle{1.24\left(\frac T{10\, {\rm keV}}\right)^{1/2}
    (1+z_{\rm coll})^{-3/2}\left(\frac{\mu}{0.59}\right)^{-1/2}h^{-1}\, {\rm Mpc}},
\end{eqnarray}
and
\begin{eqnarray}
\label{M200_our}
M_{200}(T)
&=&\displaystyle{1.55\times10^{15}\left(\frac T{10\, {\rm keV}}\right)^{3/2}
    (1+z_{\rm coll})^{-3/2}\left(\frac{\mu}{0.59}\right)^{-3/2}h^{-1}\, M_\odot},\\
\label{r200_our}
r_{200}(T)
&=&\displaystyle{1.88\left(\frac T{10\, {\rm keV}}\right)^{1/2}
    (1+z_{\rm coll})^{-3/2}\left(\frac{\mu}{0.59}\right)^{-1/2}h^{-1}\, {\rm Mpc}}.
\end{eqnarray}

Adopting the fully ionized gas value $\mu=0.59$ and the value of 
$z_{\rm coll}=0$
appropriate for a direct comparison with the simulation results of 
EMN96, equations (\ref{M500_our}), (\ref{r500_our}), (\ref{M200_our}), and
(\ref{r200_our}) above become
\begin{eqnarray}
\label{M500_our2}
M_{500}(T)
&=&\displaystyle{1.11\times10^{15}\left(\frac T{10\, {\rm keV}}\right)^{3/2}
		h^{-1}\, M_\odot},\\
\label{r500_our2}
r_{500}(T)
&=&\displaystyle{1.24\left(\frac T{10\, {\rm keV}}\right)^{1/2}
		h^{-1}\, {\rm Mpc}}.
\end{eqnarray}
and
\begin{eqnarray}
\label{M200_our2}
M_{200}(T)
&=&\displaystyle{1.55\times10^{15}\left(\frac T{10\, {\rm keV}}\right)^{3/2}
		h^{-1}\, M_\odot},\\
\label{r200_our2}
r_{200}(T)
&=&\displaystyle{1.88\left(\frac T{10\, {\rm keV}}\right)^{1/2}
		h^{-1}\, {\rm Mpc}}.
\end{eqnarray}
These are in remarkably close agreement with the empirical relations of 
EMN96 in equations (\ref{M500}), (\ref{r500}), (\ref{M200}), and (\ref{r200}),
differing from them in radius by less than 2\% and in mass by less than 
7\%, well within the error bars of those empirical relations, an excellent 
check on our TIS solution. {\it In that respect, our TIS solution, based upon
the simplifying assumptions of spherical symmetry, isothermality and 
hydrostatic equilibrium, provides an approximate analytical derivation 
of the numerically calibrated scaling laws!}
This close agreement between the predictions of our analytical TIS 
solution and the numerical 
simulation results which arise as the dynamical outcome of more complicated
initial conditions suggests that our simplifying assumptions are reasonable.

 A further
comparison with the numerical simulation results of EMN96
for the cluster gas is possible in which we compare the coefficients of the 
radius -- temperature relation at different radii, corresponding to a range of 
density contrasts $X$, derived by us according to equation (\ref{implicit}) 
above, with the numerical values reported in Table 5 of EMN96. EMN96 fit the 
simulation results for $z=0.04$ (divided by $(1+z)^{-3/2}$ to
yield the scaling law at $z=0$) and $h=0.5$ to a relation of the form
\begin{equation}
\label{rX}
r_X(T)=r_{10}(X)\displaystyle{\left(\frac T{10\,{\rm keV}}\right)^{1/2}}.
\end{equation}
A comparison of the analytical TIS predictions for $r_{10}(X)$ in this case,
with the tabulated values of EMN96 at $X=200, 250, 500, 1000$, and $2500$ is 
shown in Table~\ref{r-T}.
\begin{table*}
\centering
\begin{minipage}{140mm}
\caption{Comparison of the minimum-energy TIS solution 
and X-ray cluster simulation results for the radius-temperature relation.}
\label{r-T} 
\begin{tabular}{@{}llllll}
$X=\bar{\rho}/\rho_b$&$\zeta_X$&$\tilde{M}_X$&
	$[r_{10}(X)]_{EMN96}$\footnote{
Simulation results are from EMN96, Table 5, p. 505, with the value for $X=200$
as quoted in  Arnaud and Evrard (1998).}
	&$[r_{10}(X)]_{TIS}$ & $(TIS-EMN96)/EMN96$ \\ \hline
200..... & 24.20 &52.65 &3.7  &3.759 &0.016\\
250..... & 21.88 &48.59 &3.37 &3.398 &0.008\\
500..... & 15.96 &37.74 &2.48 &2.479 &0.0006\\
1000...  & 11.56 &28.69 &1.79 &1.796 &0.003\\
2500...  &  7.32 &18.21 &1.11 &1.137 &0.024\\ \hline
\end{tabular}
\end{minipage}
\end{table*}
The TIS and simulation values are once again in remarkably close agreement,
the same to within $2.5$ per cent at all radii, spanning over an order of 
magnitude of density contrast, including the canonical value of $X=200$ so 
widely used when cosmological simulations are analyzed in comparison with 
observations of X-ray clusters for a wide range of purposes.

Recently, Hjorth, Oukbir, and van Kampen (1998) have provided a direct 
comparison between observations of X-ray clusters and the theoretical 
mass-temperature virial relation above, which EMN96 report as a good fit to 
simulation numerical results and which we have derived from our TIS solution.
Hjorth {et al.} (1998) calibrate this relation using observations of eight
clusters in which the temperatures $T$ are determined from recent X-ray data
from the ASCA satellite while the cluster masses are determined independently
by gravitational lensing measurements. They find that the EMN96 
mass-temperature relation is in good agreement with their 
observationally-calibrated relation (i.e. within the observational 
uncertainties of roughly few$\times$10\%). 
In that sense, we can also claim good agreement between our analytical 
TIS model  mass-temperature relation and  these observations.

Regarding the numerically-calibrated radius-temperature relation of EMN96 in 
equation (\ref{r500}), which we have reproduced exactly here by our
analytically derived 
TIS solution, Mohr and Evrard (1997) have shown that observations of
nearby X-ray clusters support this relation. Once again, the analytical 
prediction of the radius-temperature relation by the TIS solution presented 
here is, therefore, consistent with the observations of X-ray clusters. 

In short, for all its simplicity, the TIS solution derived here appears to 
predict
the virial temperature and the integrated mass distribution of the X-ray 
clusters
formed in the CDM model, according to detailed, 3D,  numerical simulations of
the latter, remarkably well. It also matches the mass-temperature and
radius-temperature virial relations observed for nearby X-ray clusters.
We are encouraged by this to think that the minimum 
energy TIS solution presented here will be quite useful as a tool in 
semi-analytical models of galaxy and cluster formation and as a guide for 
understanding both the detailed numerical simulation results and observations.
Further study 
of the dynamical origin and stability of this equilibrium solution and of the 
sensitivity of the relaxation process which leads to this equilibrium state 
to initial 
conditions is warranted. In addition, it may be useful to generalize the 
results presented here to other cosmological background models, such as an 
open, matter-dominated model, or a flat model with a nonzero cosmological 
constant, in order to describe objects that form at late enough 
times in such models that departures from the Einstein - de~Sitter model may 
arise. We will address this issue in a future paper. We expect that,
for an open, matter dominated model, the structure of the postcollapse 
equilibrium object will be the same as that derived here for the Einstein - 
de~Sitter case, when expressed in dimensionless units and when referred to
the quantities at the maximum expansion epoch of the parent tophat. In other 
words, the density profile will still be as plotted in Figure~3, where the core
and truncation radii are expressed in units of $r_m$ and the density is in 
units of $\rho_{SUS}$. In that case, since the parent top-hat in an open,
matter-dominated background universe evolves at a different rate relative to
the background universe as compared to one in an Einstein-de~Sitter model,
$r_m$ and $\rho_{SUS}$ will be different.

\section*{Acknowledgments}
This research was supported in part by NASA grants NAG5-2785, NAG5-7363, and
NAG5-7821
and NSF grant ASC 9504046. PRS is grateful to his hosts at The Institute of
Astronomy, UNAM, Mexico City, for their hospitality during an early stage
of this work and for the 1997 National Chair of Excellence at UNAM which he 
received from Mexico's CONACyT. In addition, we are grateful for the 
hospitality of the Aspen Center for Physics and its Summer Astrophysics
Workshop in June 1998 during the completion of this work. We thank our 
referee Vincent Eke for a careful reading of the paper and his helpful 
comments.

\appendix
\section{An Accurate Analytical Fitting Formula for the Density profile of a
nonsingular Isothermal Sphere}  
The isothermal sphere solution can be approximated quite accurately by the 
simple analytical function
\begin{equation}
\label{rho-analyt}
\tilde{\rho}=\displaystyle{\frac{A}{a^2+\zeta^2}
    -\frac{B}{b^2+\zeta^2}},
\end{equation}
as suggested by NL97. In what follows, we adopt an approach similar to that of
NL97 to solve for the best values of the constants in equation 
(\ref{rho-analyt}), different from those in NL97, so as to provide a better
fit over the full range of radii required to describe the TIS solution derived
in this paper. We note that,
while the final result of NL97 in the body of their paper is correct, 
several formulae (specifically equations (A9) and (A11)) in their 
Appendix A2 are incorrect and thus the method should be exercised with care. 

\begin{figure}
\centerline{\psfig{figure=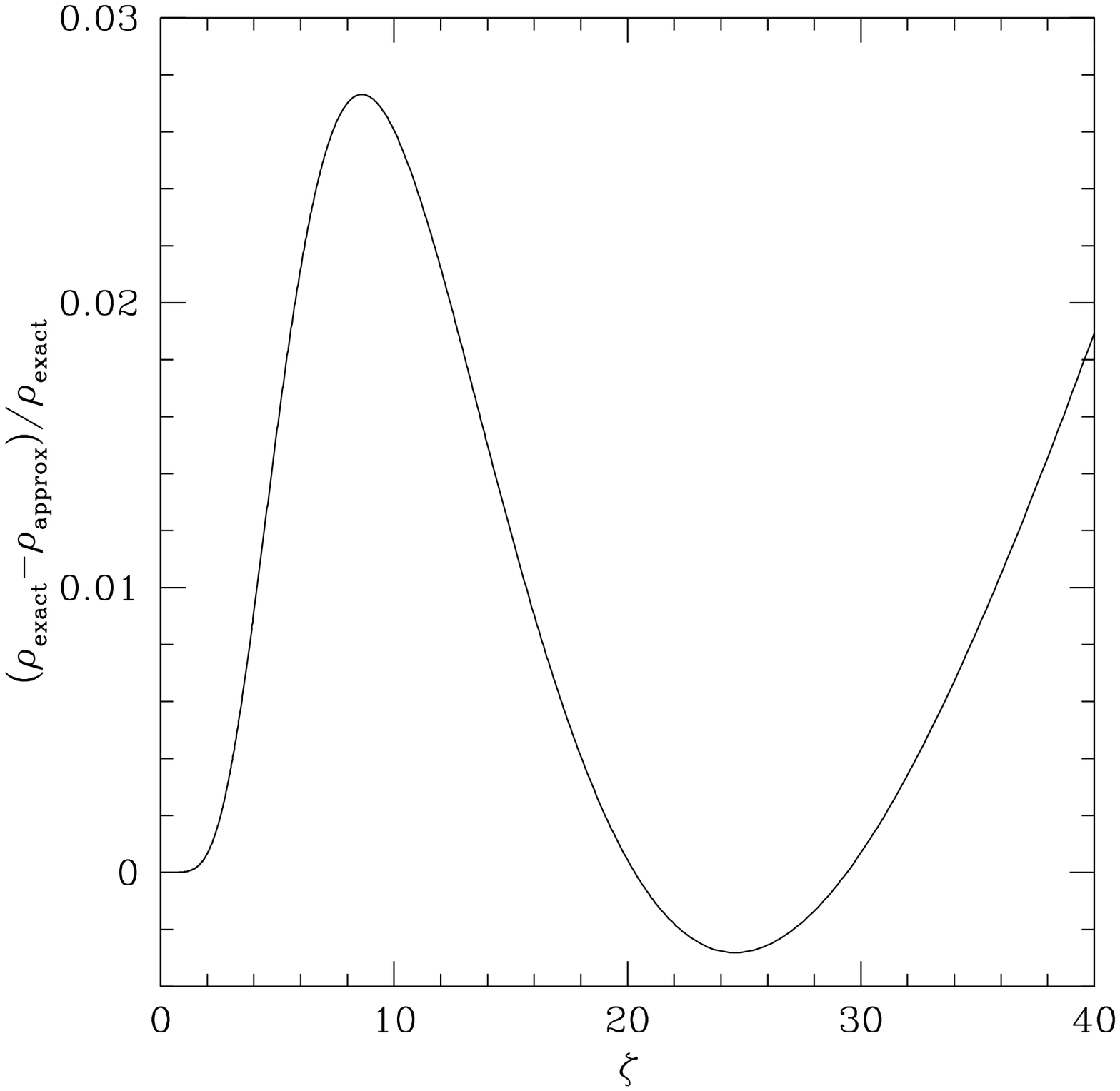,height=3in,width=3in}}
\caption{Fractional error $(\rho_{exact}-\rho_{approx})/\rho_{exact}$ of the 
approximation of equations (\ref{rho-analyt}) and (\ref{analyt_our}).}
\label{error}
\end{figure}

The solution of equation (\ref{nondim_sph}) has a Taylor expansion at $\zeta=0$:
\begin{equation}
\label{Taylor}
\tilde{\rho}=1-\frac16\zeta^2+\frac1{45}\zeta^4-\frac{61}{22680}\zeta^6
	+\frac{629}{2041200}\zeta^8+O(\zeta^{10}).
\end{equation}
Depending on our requirements on the solution, we can find different sets of
parameters for the best fit. If, for example, we require that
the first three terms of the Taylor series of equation (\ref{rho-analyt}) 
at $\zeta=0$ are the same as in equation (\ref{Taylor}) and that $\tilde{\rho}$
has the right behaviour at infinity,
\begin{equation}
\tilde{\rho}\rightarrow 2/\zeta^2\,\, {\mbox as }\qquad \zeta\rightarrow\infty,
\end{equation}
then we obtain the simple approximation of equation (\ref{Taylor}) with the 
following parameters
\begin{equation}
(A,a^2,B,b^2)_{NL,1}=(50,10,48,12).
\end{equation} 
This approximation is fairly accurate for small radii (fractional error less 
that
2 per cent for $\zeta<5$) and for very large radii (error $<2$ per cent for 
$\zeta>100$). 
However, in the intermediate interval, which we are most interested in, the
fractional error of this approximation can be as large as 24 per cent. 
To remedy this problem, NL97 offers the following 
choice of parameters:
\begin{equation}
(A,a^2,B,b^2)_{NL,2}=(44.34,10.16,42.61,12.66). 
\end{equation}
This latter approximation is an extremely good fit 
up to $\zeta=10$, but deteriorates after that, having fractional errors of 
5-8 per cent 
for $\zeta=15-40$. Hence, we have used their method, instead,  to find an 
approximation which better suits our purposes, by requiring the first three 
terms of the Taylor series to match those in equation (\ref{Taylor}),
and by imposing an exact match at $\zeta=29.4$, which produces the 
result quoted in our equation (\ref{rho-analyt_sph}) in the main text, namely
\begin{equation}
\label{analyt_our}
(A,a^2,B,b^2)_{TIS}=(21.38,9.08,19.81,14.62).
\end{equation}
While we cannot guarantee that this is the best possible
approximation of this type, it is accurate to within 3 per cent for $\zeta<40$
(Fig~\ref{error}),
while for most of the interval the error is much smaller, which should be 
sufficient for many 
purposes where using a simple functional form is more important than 
obtaining a high precision.
\end{document}